\documentclass[%
reprint,
superscriptaddress,
nofootinbib,
twocolumns,
notitlepage,
amsmath,amssymb,
aps,
pra
]{revtex4-2}

\usepackage{graphicx}
\usepackage{dcolumn}
\usepackage{bm}
\usepackage{tablefootnote}
\usepackage{multirow}
\usepackage{hhline}
\usepackage{amsthm}
\usepackage{amsmath}
\usepackage{amssymb}
\usepackage{graphicx}
\usepackage{url}
\usepackage{float}
\usepackage{bm}
\usepackage{ifthen}
\usepackage[usenames,dvipsnames]{color}
\usepackage{mathrsfs}
\usepackage[colorlinks=true,citecolor=blue,urlcolor=black]{hyperref}
\usepackage{float}
\usepackage{subfigure}
\usepackage{enumitem}
\usepackage{color}
\usepackage{setspace}
\usepackage{braket}
\usepackage{comment}
\usepackage{lipsum}

\newcommand{\be}{\begin{equation}}
	\newcommand{\ee}{\end{equation}}

\begin{document}

	\title{Numerical Security Proof for Decoy-State BB84 and Measurement-Device-Independent QKD Resistant against Large Basis Misalignment}
	
	\author{Wenyuan Wang}
	\affiliation{Institute for Quantum Computing and Department of Physics and Astronomy, University of Waterloo, Waterloo, Ontario, Canada N2L 3G1}
	
	\author{Norbert L\"utkenhaus}
	\affiliation{Institute for Quantum Computing and Department of Physics and Astronomy, University of Waterloo, Waterloo, Ontario, Canada N2L 3G1}
	
	\begin{abstract}
		In this work, we incorporate decoy-state analysis into a well-established numerical framework for key rate calculation, and apply the numerical framework to decoy-state BB84 and measurement-device-independent (MDI) QKD protocols as examples. Additionally, we combine with these decoy-state protocols what is called ``fine-grained statistics", which is a variation of existing QKD protocols that makes use of originally discarded data to get a better key rate. We show that such variations can grant protocols resilience against any unknown and slowly changing rotation along one axis, similar to reference-frame-independent QKD, but without the need for encoding physically in an additional rotation-invariant basis. Such an analysis can easily be applied to existing systems, or even data already recorded in previous experiments, to gain significantly higher key rate when considerable misalignment is present, extending the maximum distance for BB84 and MDI-QKD and reducing the need for manual alignment in an experiment.
	\end{abstract}
	
	\date{\today}
	\maketitle
	
	\section{Background}
	
	Quantum Key Distribution (QKD) \cite{BB84} can provide information-theoretic security between two communicating parties, Alice and Bob. Until recently, a new proof had to be derived for each new type of protocol or new side channel considered. 
	
	In Refs. \cite{numerical1, numerical2}, a novel numerical framework has been proposed that can take in a universal set of descriptions for a protocol and the simulated/experimental data from a quantum channel, and numerically bound the key rate with a streamlined algorithm. Such a numerical framework has the great advantage of being able to use a common algorithm to calculate the key rate for various protocols, thus avoiding specialized approaches that work only for each particular setting. The streamlined approach also makes it easier to apply finite-size analysis to protocols. The numerical framework has been successfully applied to various protocols such as BB84, measurement-device-independent (MDI) QKD, three-state protocol, discrete-modulated continuous variable (CV) QKD, and side channels such as unbalanced phase-encoding and detector efficiency mismatch in BB84 \cite{numerical1,numerical2,numerical_CV,numerical_blockdiag}. The finite-size analysis for the framework is demonstrated in Ref. \cite{numerical_finite}.
	
	One limitation in the previous works is that the numerical approach has not yet been combined with the established technique of decoy states. Most of the applications above consider only single photon sources, except for Ref. \cite{numerical_blockdiag} that uses an infinite number of decoys that perfectly yield the statistics from each photon number state, and Ref. \cite{numerical_CV} that performs discrete-modulated CV QKD with a finite number of non-phase-randomized coherent states. For many discrete-variable QKD protocols, phase-randomized weak coherent pulse (WCP) sources are often used, and decoy-state analysis is often used to bound the single-photon contributions. A small number (e.g. two to three) of decoy intensity settings are often used \cite{decoy_practical,MDIPrac}.
	
	In this work, we apply the numerical framework to two important protocols, BB84 \cite{BB84} and MDI-QKD \cite{MDI}, in the case of using WCP sources and finite number of decoy states. The BB84 protocol is widely used in both fibre-based and free-space QKD experiments \cite{QKDfibre,QKDfreespace}, and the MDI-QKD protocol enables immunity against all hacking attacks on the detector, which not only sees multiple experimental implementations, but has recently been demonstrated over free-space too \cite{MDIfibre,MDIfreespace}. We show that we can incorporate decoy states into the numerical proof by simply considering it as a ``wrapper" that preprocesses the detection data and estimates the single-photon contributions to the observables. This allows us to reduce the problem to a finite-dimensional one, which can then be fed to the numerical solver, making it useful with practical scenarios using WCP sources. Importantly, the wrapper approach we describe is a rather general tool: in principle, any protocol that can be described in the numerical framework can be uplifted to a decoy-state protocol. A protocol can be described in a smaller Hilbert space (e.g. single photon based qubits), which can be mixed with signals from other larger spaces, while the decoy-state wrapper can preprocess the detection data and single out the statistics in the Hilbert space of interest (without modifying the descriptions of the original protocol).
	
	The second aspect of our work is the use of what is called ``fine-grained statistics" in the numerical framework. This is a technique where Alice and Bob use the full set of detector click data, including the cross-basis events where they chose different bases (which are often discarded) in the security analysis, to better characterize the channel and get higher key rate. This is rather natural for a numerical approach, because it treats all observed data as constraints in an optimization problem, and adding more constraints to the same problem simply helps us get a tighter bound on the key rate, without requiring any modifications to the security proof framework. The technique has previously been applied to BB84 with the numerical framework and combined with finite-size analysis in Ref. \cite{numerical_finite}. This idea of using cross-basis/discarded data has also been discussed in many previous works, such as e.g. Refs. \cite{finegrain,finegrain2, RFI,RFIMDI}, notably in the application of reference-frame-independent (RFI) QKD \cite{RFI}. In a way, the usage of fine-grain statistics defines a set of modified QKD protocols based on given QKD protocols, where full data instead of sifted data is used for security analysis. 
	
	In this work we apply fine-grained statistics to decoy-state BB84 and MDI-QKD, and show that the modified protocols using fine-grained statistics enjoy a much higher resilience against basis misalignment, because they are able to make use of more information to effectively characterize the channel. \footnote{Throughout the text we will mainly consider polarization-encoding protocols through optical fibre (where we can model misalignment as a unitary rotation to an angle), but by a broader definition, misalignment is also present in other degrees of freedom such as phase encoding or time-bin phase encoding.} 
	
	Note that resilience against misalignment is an important advantage, because misalignment errors pose a major challenge to the experimental implementation of both BB84 and MDI-QKD, whether implemented via fibre or free-space channels. It is especially severe for fibre-based MDI-QKD as it relies on two-photon interference which is sensitive to the misalignment between two incoming signals, and it is also a problem for free-space QKD as there could be limited time for alignment and communication. To mitigate misalignment, one typically implements automatic feedback of polarization, which increases the complexity of the system and gets more costly the faster and the more precisely it has to work. Therefore, gaining resilience against misalignment on the protocol level is highly desirable.

 	Like the limitations of RFI-QKD, our approach does still have some limitations in that it requires that (1) the rotation angle is \textit{slowly drifting} (and can be assumed to be constant throughout one experiment session), and (2) the rotation is only along \textit{a fixed axis} perpendicular to the plane on the Poincar\'e sphere on which testing signals are sent, such as the X-Z plane or the X-Y plane, depending on which bases are used for testing.
	
	As mentioned above, the RFI protocols \cite{RFI,RFIMDI} have been previously proposed as a solution to large misalignment. Such protocols make use of two bases X and Y to encode testing data, and use a third basis Z to encode the key generating data. The full data (including cross-basis clicks) between X and Y bases are used to perform a tomography-like analysis, which estimates the virtual phase error for the Z basis signals. In a way, this is a special case of applying the fine-grained statistics idea, except that the data is not directly incorporated as constraints into the analysis, but used to first construct a constant that is invariant under misalignment, and which is later used to bound the phase-error rate. The RFI approach has also been applied to MDI-QKD to form RFI-MDI-QKD protocols. The key point of RFI-QKD is that it is usually easier to maintain good alignment of one basis, e.g. the Z basis (which can correspond to the circular polarization for polarization encoding, or the time-bin basis for time-bin phase encoding), while the other two basis can be allowed to slowly drift, resulting in a unitary rotation of an unknown angle along only one plane. A limitation to RFI protocols is that they require the use of an additional encoding basis, which requires additional physical modification to the systems to work and increases the complexity. 

	In comparison, we use the fine-grained statistics directly as constraints in the numerical framework. We also do not require any physical modification to the QKD protocol it is based on (as we are only making use of data that are supposed to be discarded), which means we can use an existing experimental setup, or even potentially apply the analysis to experimental data acquired from previously completed experiments and readily obtain a higher key rate due to a refined analysis.
	
	In this manuscript, we will introduce our methodology in Sec. II where we briefly recapitulate the numerical framework, introduce how we combine decoy state analysis with the framework, and introduce how to use the fine-grained statistics. In Sec. III we introduce the channel model we use to simulate the statistics. We will then present our simulation results in Sec. IV and conclude in Sec. V.

	\section{Method}
	
	In this section we introduce the methodology we use to bound the secure key rate for decoy-state BB84 and MDI-QKD. We will briefly recapitulate the numerical framework in Ref. \cite{numerical2}, the decoy-state method \cite{decoy1,decoy2,decoy3}, the concept of using fine-grained statistics, and the channel model we use for our simulation.
	
	\subsection{Numerical Framework}
	
	Here in this subsection, we first give a high-level description of the numerical framework we use to calculate the key rate. The details of the numerical framework can be found in Ref. \cite{numerical2}. 
	A simple step-by-step description for a protocol has a typical structure as follows:	
	\begin{enumerate}
		\item {\bf Quantum Phase} In an entanglement-based scheme, a source sends quantum signals to Alice and Bob, who measure the incoming signals . In a prepare-and-measure scheme, Alice prepares a signal and sends it to Bob, who measures it; 
		\item {\bf Testing}  Alice and Bob announce their detailed measurement outcome (and choice of prepared signal) for a subset that they selected at random;
		\item {\bf  Classical Communication Phase} Alice and Bob perform some block-wise post-processing of data, for example by announcing some information, such as basis announcements, publicly.
		\item {\bf Key Map} One party, say Alice, performs a key map, which is a function of her local data and the information exchanged in the block-wise post-processing, to obtain a data set from which a secret key will be distilled with the next two steps. Alice applies the key map on the state to obtain the raw keys. Note that only Alice implements the key map (and the result is not changed by the next error correction step).
		\item {\bf Error Correction} Alice and Bob perform error correction by exchanging classical information. They keep track of the resulting information leakage from the key map step.
		\item {\bf Privacy Amplification} Alice and Bob perform privacy amplification, whose parameters are set by the security analysis, based on the observation during the testing phase and the leaked information during the error correction step.
	\end{enumerate}

	The key idea in the numerical framework that provides the security analysis mentioned in the Privacy Amplification step above can be fully described by a few formal elements. Note that the theoretical description of the prepare-and-measure protocols can be unified using the concept of a thought set-up via the source-replacement scheme {\cite{sourcereplacement1,sourcereplacement2,sourcereplacement3,sourcereplacement4}. 
	
	The physical set-up of QKD devices is reflected by
	\begin{itemize}		
		\item   \textit{measurements} performed by Alice and Bob which are described by general Positive Operator Valued Measures (POVMs), which are formed by POVM elements that are represented by matrices $\{\Gamma_k\}$. Here the POVM elements can be ``fine-grained" (containing all raw statistics) or ``coarse-grained" (combined into groups), a process we will describe in more detail in Sec. II.C;
		
		\item \textit{expectation values} of the measurements (numerical values $\{\gamma_k\}$) which corresponds to the POVMs described for the protocol. These expectation values are obtained during the Testing Phase of the QKD protocol. (Since we are dealing with the asymptotic case, the frequency of observed events is described by the corresponding quantum mechanical probabilities.) 
	\end{itemize}

	The POVMs $\{\Gamma_k\}$ and observations $\{\gamma_k\}$ (which asymptotically correspond to the expectations of POVMs) form constraints that bound the possible input states $\rho$ within a domain $S$ satisfying $Tr(\Gamma_k \rho) = \gamma_k$.\\
	
	The post-processing phase is represented by 
		\begin{itemize}	
		\item the \textit{Kraus operators} $\{K_i\}$, which represent measurements, public announcements and post-selection by Alice and Bob (such as basis choice). The process can be viewed of as a quantum channel acting on the received quantum state that Alice and Bob share, and it creates also classical registers, embedded in a quantum states, corresponding to the announcements and the processing. 
		
		
		\item the \textit{key map operators} $\{Z_j\}$: After announcements and post-selection, Alice performs the key map making use of locally available information and the publicly communicated information. This can be thought of as a measurement on the quantum registers holding the the relevant information. This measurement is mathematically represented by a pinching quantum channel, corresponding to a partial dephasing along blocks that are characterized by the key map measurement results. The operators $\{Z_j\}$ are projects onto the subspaces corresponding to these blocks.	
	\end{itemize}

	Based on the Renner framework \cite{renner}, and using the formulation in Refs. \cite{numerical1,numerical2}, the key rate can be lower-bounded by
	
	\begin{equation}
		R = \text{min}_{\rho \in S} \: f(\rho) - p_{\text{pass}} \times \text{leak}^{\text{EC}}_{\text{obs}}
	\end{equation}

	\noindent where $\text{leak}^{\text{EC}}_{\text{obs}}$ is the error-correction leakage and $p_{\text{pass}}$ is the probability of a signal being detected (the yield) and passing the basis sifting. The domain S that bounds possible values of $\rho$ is
	
	\begin{equation}
		S(\vec{\gamma})= \{\rho \in H_+ \:|\: Tr(\Gamma_k \:  \rho) = \gamma_k, \:\forall k \}.
	\end{equation}

	Note that for the privacy amplification term, $p_{\text{pass}}$ is already included in $f(\rho)$. This is because information such as the channel loss and basis choice probabilities are already included in the statistics $\gamma_k$, which subsequently limit $\rho$ to only those that have a smaller $f(\rho)$ that corresponds to the correct amount of channel loss and sifting.
	
	Here $f(\rho)$ is solely determined by the Kraus operators $\{K_i\}$ and key map operators $\{Z_j\}$:

	\begin{equation}
		\begin{aligned}
			f(\rho) &= D(\mathcal{G}(\rho) || \mathcal{Z(\mathcal{G}(\rho))} ) \\
		\end{aligned}
	\end{equation}

	\noindent where
	
	\begin{equation}
		\begin{aligned}
			\mathcal{G}(\rho) &= \sum_i  \: K_i  \: \rho  \: K_i^\dagger, \\
			\mathcal{Z(\mathcal{G}(\rho))} &= \sum_j  \: Z_j  \: \mathcal{G}(\rho)  \: Z_j \\
		\end{aligned}
	\end{equation}

	\noindent and $D(X||Y)=Tr(X \: \text{log}X) - Tr(X \: \text{log}Y)$ is the quantum relative entropy (where $\text{log}$ is the matrix logarithm).

	The problem of minimizing $f(\rho)$ can be broken down to a two-step approach: in a first step one  finds a near-optimal $\rho'$  and in a second step one  solves a linearized dual problem at $\rho'$ to find the rigerous lower bound. The details can be found in Ref. \cite{numerical2}.

	In this paper we also compare results to existing works using quantum error correction code approach (using Shor-Preskill \cite{ShorPreskill} key rate):
	
	\begin{equation}
		R = 1-h_2(e_1^X) - h_2(e_1^Z).
	\end{equation}

	For BB84 using weak coherent pulse source and decoy-states \cite{decoy1,decoy2,decoy3}, this becomes the key rate from the single-photon proportion \footnote{Here for BB84 and below for MDI-QKD, we assume only the Z basis is used for generating keys, while in principle both basis can be used; alternatively, in the asymptotic case it is also possible to assume $p_Z \approx 1$ such that there is no loss of key rate from sifting.}:
	
	\begin{equation}
		R = p_Z^2  \: p_{1}  \: Y_{1}  \: [1-h_2(e_{1})] - p_Z^2 \:  Q_{\mu} \:  h_2(E_\mu))
	\end{equation}
	
	\noindent where $Q_{\mu}$ is the signal state detection probability (gain), $p_{1}$ is the probability of a single-photon being sent (following a Poissonian distribution for WCP source), and $Y_{1},e_{1}$ are the single photon statistics (yield and error rate) estimated from decoy state analysis, which will be discussed in more detail in the next subsection. Or, for MDI-QKD based on the portion of single-photon pairs from WCP sources \cite{MDI,MDIPrac}, the key rate is
	
	\begin{equation}
		R = p_Z^2  \: p_{11}  \: Y_{11}^Z  \: [1-h_2(e_{11}^X)] - p_Z^2  \: Q_{\mu\mu}^Z  \: h_2(E_{\mu\mu}^Z)
	\end{equation}
	
	\noindent where $Q_{\mu\mu}^Z$ is the signal state detection probability (gain), $p_{11}$ is the probability of single-photons begin sent by both Alice and Bob (i.e. $p_1^2$ if they use the same intensity), and $Y_{11}^Z,e_{11}^X$ are the single photon statistics estimated from decoy state analysis.
		
	\subsection{Decoy-State Analysis}
The BB84 and MDI-QKD  protocols have the best performance if the signals were all single-photons. However, single photon sources are technologically less mature, and are more challenging to protect against side-channel attacks, as side-channel protection typically introduces loss elements in the optical path. The associated loss can be easily compensated by an amplitude adjustment for laser pulses, but will invariably result in a performance degradation for single-photon source based schemes. Despite that, there has been more recent progresses made, e.g. using quantum dot single photon source for QKD \cite{singlephoton}. Meanwhile, a mature and widely used source is the phase-randomized weak coherent pulses (WCP) source. As observed in Refs. \cite{dephase1, dephase2}, dephasing the coherent source allows the coherent state to be described as a mixture of Fock states (where Eve's photon-number-splitting attack does not affect the single photon component). The dephasing method still has limitations, though, such as in maintaining the phase-randomization for sources at high repetition rates. For a dephased WCP source, the photon numbers follow a Poissonian probability distribution (with mean photon number $\mu_i$):

	\begin{equation}
	p_{\mu_i}(n) = {{\mu_i}^n \over n!}  \: e^{-{\mu_i}}.
	\end{equation}

	From the experiments we will only be able to observe the statistics from the WCP source (which is a mixture of statistics from all photon number states). If we denote an observable of interest as $\gamma$, the contribution from n-photon states as $\gamma_n$, and the overall observed statistics from the WCP source as $\gamma_{\mu_i}$, we can write
	
	\begin{equation}
		\begin{aligned}
			\gamma_{\mu_i} &= \sum p_{\mu_i}(n)  \: \gamma_n.
		\end{aligned}
	\end{equation}

	The decoy-state analysis proposed in Refs. \cite{decoy1,decoy2,decoy3} is a classical post-processing technique that estimates the statistics of single photons (i.e. $\gamma_1$), by combining the data from multiple different intensities (hence different distributions) and upper/lower bounding the single photon contribution among the data. The key assumption for decoy-state analysis to hold is that \textit{for any given photon number state being sent, the Eavesdropper cannot tell which decoy intensity choice it came from}. This is a valid assumption because the sampling of photon number state from the Poissonian distribution is a Markov process, i.e. the photon number state does not contain information of the intensity of the source it came from.
	
	
	The process of estimating single photon contributions can be formulated numerically as a linear programming problem \cite{decoy_numerical}. For instance, the linear program constraints for BB84 can be written as:
	
	\begin{equation}
		\begin{aligned}
			\gamma_{\mu_i} &\leq \sum_{n\leq N}  \: p_{\mu}(n)  \: \gamma_n  + (1-\sum_{n\leq N}  \: p_{\mu}(n)),\\
			\gamma_{\mu_i} &\geq \sum_{n\leq N}  \: p_{\mu}(n)  \: \gamma_n.\\
		\end{aligned}
	\end{equation}

	\noindent where $\{\gamma_n\}$ are the variables and constants $\{\gamma_{\mu_i}\}$ are the statistics from the set of decoy state intensities used. Also, to reduce the infinite number of variables to finite, here we need to apply a photon number cutoff $N$, where variables with $n>N$ are upper (lower) bounded by 1 (0). Similar applies to MDI-QKD, where photon numbers for Alice and Bob each satisfy the cutoff condition. These techniques for performing the decoy-state analysis and solving for single-photon statistics have been well described in previous works, such as Refs. \cite{decoyreview,decoy_practical, decoy_numerical,MDIPrac}. More details on the formulation of the linear programming model can be found in Appendix B. 

	In practice, there could be a set of observables of interest from Bob's set of POVMs, which we can denote as a vector $\vec{\gamma}$ whose components are observables $\gamma_k$. Using decoy-state analysis, we can obtain the upper and lower bound for the single-photon contributions for each observable. In other words, we find a bound $G$ such that the single photon statistics satisfy $\vec{\gamma}_1 \in G$, where
	
	\begin{equation}
		G = \{\vec{\gamma}_1 \in R^K \:|\: \gamma_{1,k} \in [\gamma_{1,k}^L,\gamma_{1,k}^U], \:\forall k \}\\
	\end{equation}

	\noindent is defined as the set where each component $\gamma_{1,k}$ of $\vec{\gamma}_1$ satisfies the bounds $\gamma_{1,k}^L,\gamma_{1,k}^U$ obtained from decoy-state analysis (assuming there are a total of $K$ components, i.e. observables).\\
	
	Now, with the bounds on single-photon contributions known from decoy-state analysis, we discuss how to incorporate it into the numerical framework. Importantly, for our numerical framework here, we will show that the decoy-state analysis can be considered as classical preprocessing of observed data acquired from the channel, before the estimated single-photon contributions are used as a ``pseudo-observable" and used to bound the density matrix and minimize key rate.
	
	Below we include a more rigorous formulation for integrating the decoy-state analysis into the numerical framework. Here we first recapitulate the results of Ref. \cite{numerical_blockdiag}, which considers a phase-randomized coherent source and an infinite number of decoy settings. It is shown that Alice's source can be formulated as a local qubit entangled with a ``shield system" (representing the photon number sent) and the signal states (which are Fock states containing encoding information) being sent, such that, when traced over the shield system, the output signal state is in a mixed state:
	
	\begin{equation}
		\ket{\Psi}_{AA_sA'} = \sum_{x} \sqrt{p_x} \ket{x}_A \otimes \sum_{n} \sqrt{p_n} \ket{n}_{A_s} \ket{s_n^x}_{A'},
	\end{equation}
	
	\noindent where $p_x$ is the probability of choosing each signal state, and $p_n$ is the Poissonian distribution for sending a photon number state. The system $A'$ is sent through the channel and measured by Bob. The important point is that, due to the phase randomization, the sent state is a mixture of Fock states (of which we assume the photon number is also known by Eve), and the final state $\rho_{AA_sB}$ is a \textit{block-diagonal} state where the blocks are defined by the photon number (in the shield system $A_s$):
	\begin{equation}
		\rho_{AA_sB} = \sum_{n} p_n \ket{n}\bra{n}_{A_s} \otimes \rho_{AB}^{(n)} \; .
	\end{equation}
Here  $\rho_{AB}^{(n)}$ is the shared density matrix for systems $A$ and $B$, conditional to $n$ photons being sent. From Ref. \cite{numerical_blockdiag}, the key rate can be written as:
	
	
	\begin{equation}
	\begin{aligned}
		R &\geq \left[ \sum_n \: p_n \: \text{min}_{\rho_{AB}^{(n)} \in S_n} \: f\left(\rho_{AB}^{(n)}\right) \right] - p_{\text{pass}} \times \text{leak}^{\text{EC}}_{\text{obs}},\\
	\end{aligned}
	\end{equation}
	
	\noindent where $f(\rho)=D(\mathcal{G}(\rho) || \mathcal{Z(\mathcal{G}(\rho))} )$ described the amount of extractable key during privacy amplification (aside from the error correction cost) given a density matrix $\rho$, and $p_n$ is again the Poissonian distribution. Note that, as mentioned in Sec. II.A, the detection probability (yield) and sifting factor for the privacy amplification term are already contained in $f(\rho)$, since the constraints on $\rho$ contain information about them. It is shown in Ref. \cite{numerical_blockdiag} that, due to $\rho_{AA_sB}$ being block-diagonal, we can independently calculate the privacy amplification for each conditional scenario where $n$ photons are sent, while optimizing each $\rho_{AB}^{(n)}$ independently. Note that, since $f$ is defined only based on the Kraus operators and key maps (which only involve systems $A,B$ and registers that are independent of the actual signal sent), it stays in the same form for all $\rho_{AB}^{(n)}$.\\
	
	Now, for our scenario, we are only interested in the single-photon contribution from the source. Here if we consider polarization-encoding, the multi-photon components are not going to generate key rate due to Eve's ability to perform photon-number-splitting attacks. \footnote{For protocols where multi-photons do contribute to key rate, since quantum relative entropy is non-negative (i.e. there is no negative contribution from any photon number states), we can still study the single-photon part, but the first line of Eq. 15 would be an inequality too.} We can then simply write:
	
		\begin{equation}
		\begin{aligned}
			R &= p_{\text{pass}}^{(0)} + p_1 \: \text{min}_{\rho_{AB}^{(1)} \in S_1} \: f\left(\rho_{AB}^{(1)}\right) - p_{\text{pass}} \times \text{leak}^{\text{EC}}_{\text{obs}}.\\
			&\geq p_1 \: \text{min}_{\rho_{AB}^{(1)} \in S_1} \: f\left(\rho_{AB}^{(1)}\right) - p_{\text{pass}} \times \text{leak}^{\text{EC}}_{\text{obs}}.\\
		\end{aligned}
	\end{equation}


	Note that, importantly, as pointed out by Ref. \cite{numerical_blockdiag}, as well as some earlier works \cite{zerophoton1,zerophoton2,zerophoton3} on analytical bounds for the key rate of BB84 and MDI-QKD, here the zero-photon component actually contributes to the key rate too, since Eve cannot know what local state Alice prepared at all with no photon being sent and all clicks coming from dark counts. The zero-photon contribution is simply
	\begin{equation}
		p_0 \: \text{min}_{\rho_{AB}^{(0)}\in S_0}\:f\left(\rho_{AB}^{(0)}\right)=p_{\text{pass}}^{(0)}
	\end{equation}
	 
	 \noindent where $p_{\text{pass}}^{(0)}$ is the probability of a zero-photon event being detected and passing sifting. Nevertheless, to save computational time and avoid performing an additional decoy-state analysis on zero-photon components, and to compare with known analytical formulae (since many references, such as Refs. \cite{decoy_practical,MDI} we compare to, omit the zero-photon contribution), here in the following discussions and the simulations in Sec. IV we will also omit the zero-photon term, but in principle we can always choose to use it to get even higher key rate. 
	 
	For the single-photon subspace, the bound $S_1$ is defined by single-photon statistics $\vec{\gamma}_1$ \footnote{In fact here $S_1$ also contains local constraints that characterize the source, which stay unchanged regardless of the channel and we have omitted for simplicity.}:
	
	\begin{equation}
		S_1(\vec{\gamma}_1)= \{\rho_{AB}^{(1)} \in H_+ \:|\: Tr(\Gamma_k \:  \rho_{AB}^{(1)}) = \gamma_{1,k}, \:\forall k \},
	\end{equation}
	
	For simplicity we will denote $\rho_{AB}^{(1)}$ as $\rho$ from now on.\\
	
	Since the single-photon statistics are not known when using a WCP source, we cannot accurately obtain $S_1$. However, as mentioned above, we know from decoy-state analysis the bounds for the single-photon statistics $\vec{\gamma}_1 \in G$. Each valid $\vec{\gamma}_1$ generates an individual set $S_1(\vec{\gamma}_1)$. The key rate can be written as the worst-bound over all possible $\vec{\gamma}_1$:
	
	\begin{equation}
		\begin{aligned}
			R &\geq p_1 \: \text{min}_{\vec{\gamma}_1 \in G}  \left[ \text{min}_{\rho\in S_1(\vec{\gamma}_1)} \: f(\rho) \right] - p_{\text{pass}} \times \text{leak}^{\text{EC}}_{\text{obs}}.\\
		\end{aligned}
	\end{equation}

	However, we do not need to perform a double optimization here. Instead, we can define a union set $S_1'(G)$ of all sets $S_1(\vec{\gamma}_1)$ for every $\vec{\gamma}_1 \in G$, which mathematically takes the simple form of:
	
	\begin{equation}
		S_1'(G)= \{\rho \in H_+ | \gamma_{1,k}^L \leq Tr(\Gamma_k \:  \rho) \leq \gamma_{1,k}^U, \:\forall k \},
	\end{equation}

	\noindent which we can see is simply Eq. 17 with the equality constraints replaced by inequalities (loosened bounds) obtained from decoy-state analysis. We can rewrite the final key rate as:
	
	\begin{equation}
		\begin{aligned}
			R &\geq p_1 \:  \text{min}_{\rho \in S_1'(G)} \: f(\rho) - p_{\text{pass}} \times \text{leak}^{\text{EC}}_{\text{obs}},\\
		\end{aligned}
	\end{equation}

	The decoy-state analysis functions like a ``wrapper" here: it helps us to first obtain bound $G$, which is then passed to the actual optimization as the pseudo-statistics in bounding the key rate. Note that the protocol description is the same, i.e. $f$ is independent of the decoy-state procedure, and in principle, any protocol that can be described in our numerical framework can be uplifted to a decoy-state based protocol by first applying the wrapper to obtain the bounds on desired photon number statistics and then calculating the key rate based on these pseudo-statistics.
	 
	 An additional point is that, while in this work we  only consider the single photon contribution to key rate, in principle the technique is also applicable to multi-photon contributions, since for certain protocols (such as the one shown in Ref.\cite{numerical_blockdiag}), multi-photon components might contribute to key rate. 

	For multi-photon components, surely it is possible to independently find the upper/lower bounds on each set of n-photon statistics $\vec{\gamma}_n$ such that $\vec{\gamma}_n \in G_n$, and from each $G_n$ we can separately optimize the density matrix $\rho_{AB}^{(n)}$ in the subspace for a given photon number sent and calculate the n-photon contribution to the key rate. However, in this way we are underestimating the total key rate because the statistics for multi-photons ($\vec{\gamma}_n$) are jointly bound by the same set of linear constraints, in the form of Eq. 10, and they will not simultaneously take the worst-case upper/lower bounds, such as those acquired from independently optimizing each $\vec{\gamma}_n$. However, it requires more substantial computational resources to solve a problem where all density matrices $\rho_{AB}^{(n)}$ are jointly optimized (subject to the condition that all $\vec{\gamma}_n$ jointly satisfy the linear constraints), and such a problem will be a subject of future studies.
	 
	
	\subsection{Fine-Grained Statistics}
	
	As we discussed in Sec. I, one can use the fine-grained statistics \cite{numerical_finite} to incorporate the full set of data including all cross-basis events for Alice and Bob in bounding Eve's information and acquiring the secure key rate. This is especially convenient for the numerical approach, since each detection event (the expectation for a POVM) simply corresponds to a constraint in the SDP problem. Using such cross-basis data does not affect the description of a protocol (the Kraus operators and key maps), but simply introduces more constraints to the optimization, which potentially can provide a tighter bound and a better key rate. 
	
	Below we show how fine-grained statistics can be used for BB84 and MDI-QKD. 
	
	For BB84, the format of data is shown in Table I. Here the observed statistics is a $4\times 5$ matrix, corresponding to 4 input states encoded by Alice, and 5 POVMs corresponding to the four measurement outcomes $H,V,+,-, \varnothing$ (with $\varnothing$ meaning a failure event, such as obtaining no click, or registering double clicks from detectors measuring different bases). Table I shows the format of measurement data that can be obtained from e.g. a simulation or an experiment. We denote the first subscript as the sender's encoded state and second as the receiver's POVM. Asymptotically, these entries correspond to the expectation values of POVMs.
	
	\begin{table}
		
		\begin{tabular}{|c|c|ccccc|}
			\hline
			&&\multicolumn{5}{c|}{Bob measures}\\
			\hline
			&&H&V&+&-&$\varnothing$\\ 
			\hline
			&H & ${\color{red}\gamma_{HH}}$&${\color{red}\gamma_{HV}}$&$\gamma_{H+}$&$\gamma_{H-} $&$ \gamma_{H\varnothing}$\\ 
			\multirow{2}{*}{Alice}&V & ${\color{red}\gamma_{VH}}$&${\color{red}\gamma_{VV}}$&$\gamma_{V+}$&$\gamma_{V-} $&$ \gamma_{V\varnothing}$\\ 
			\multirow{2}{*}{sends}&+ & $\gamma_{+H}$&$\gamma_{+V}$&${\color{red}\gamma_{++}}$&${\color{red}\gamma_{+-}}$&$ \gamma_{+\varnothing}$\\ 
			&- & $\gamma_{-H}$&$\gamma_{-V}$&${\color{red}\gamma_{-+}}$&${\color{red}\gamma_{--}} $&$ \gamma_{-\varnothing}$ \\
			\hline
			
		\end{tabular}
		\caption{Format of fine-grained statistics for BB84 with channel loss considered. $\varnothing$ means a failure event. The entries marked in red are the ones used in the traditional approach with gain and QBER (coarse-grained data). In our approach the full table (fine-grained data) is used.}
	\end{table}

	To use fine-grained statistics, we use each term (a total of 20 terms) as a constraint for the numerical solver, while in the traditional approach using gain and QBER (we can denote it as ``coarse-grained" statistics), we will only make use of the eight matched-basis events, marked above in red. We can denote e.g. $\gamma_{HH}+\gamma_{VV}+\gamma_{HV}+\gamma_{VH}$ as the gain and $\gamma_{HV}+\gamma_{VH}$ as the QBER (or alternatively we can simply use the eight terms as constraints). The key point is that, in the traditional approach, mismatched-basis data is discarded, while for fine-grained statistics all data is used.
	
	For MDI-QKD, the statistics is a $4\times 4\times3$ matrix, corresponding to 4 input states each for Alice and Bob, and 3 Bell state measurement outcomes $\Psi^+,\Psi^-,\varnothing$ from Charlie's detectors. Here $\Psi^+,\Psi^-$ corresponds to two types of specific detector patterns signifying successful Bell measurement outcomes for two of the four Bell states, while every other pattern is considered a failure event denoted by $\varnothing$. Each $4 \times 4$ slice for a given detector pattern for Charlie looks like Table II, which includes all possible combinations for Alice's and Bob's encoded bits. Again, only the terms marked in red are used in the coarse-grained statistics, while the terms marked in black are cross-basis clicks which are originally discarded, but now we include them into the security analysis and use these information to derive a tighter bound on Eve's information.

	\begin{table}
		
		\begin{tabular}{|c|c|cccc|}
			\hline
			&&\multicolumn{4}{c|}{Bob sends}\\
			\hline
			&&H&V&+&-\\ 
			\hline
			&H&${\color{red}\gamma_{HH}^{\Psi^+}}$&${\color{red}\gamma_{HV}^{\Psi^+}}$&$\gamma_{H+}^{\Psi^+}$&$\gamma_{H-}^{\Psi^+} $\\ \multirow{2}{*}{Alice}&V&${\color{red}\gamma_{VH}^{\Psi^+}}$&${\color{red}\gamma_{VV}^{\Psi^+}}$&$\gamma_{V+}^{\Psi^+}$&$\gamma_{V-}^{\Psi^+} $\\ \multirow{2}{*}{sends}&+&$\gamma_{+H}^{\Psi^+}$&$\gamma_{+V}^{\Psi^+}$&${\color{red}\gamma_{++}^{\Psi^+}}$&${\color{red}\gamma_{+-}^{\Psi^+}}$\\ &-&$\gamma_{-H}^{\Psi^+}$&$\gamma_{-V}^{\Psi^+}$&${\color{red}\gamma_{-+}^{\Psi^+}}$&${\color{red}\gamma_{--}^{\Psi^+}} $\\
			\hline
			
		\end{tabular}
		\caption{Format of fine-grained statistics for MDI-QKD with channel loss considered. The first subscript represents Alice and the second represents Bob, and the superscript represents Charlie's detection. There should be three outcomes for Charlie ($\Psi^+,\Psi^-,$Fail), although here to save space we have only showed the slice of data for outcome $\Psi^+$ announced by Charlie as an example. The full table should be $4\times4\times3$ in size. Again, coarse-grained data only uses the entries marked in red, while the fine-grained data incorporates the full table.}
	\end{table}

	\noindent 
	
	Once we perform decoy-state analysis, and incorporate all cross-basis events, we can expect to observe an increase in the key rate due to reduced requirement for privacy amplification, as we have a better knowledge of the misalignment, i.e. rotation operation, in the channel (similar to the fashion of performing a tomography).
	
	\subsection{Fine-Grained Statistics in Key Generation}
	
	A problem to note is that, although incorporating the fine-grained statistics into the security analysis reduces the effect of misalignment on privacy amplification, the above technique does not change the amount of classical leakage caused by error correction, which directly depends on the misalignment in the signal state. 
	
	In principle, it is still possible to make use of fine-grained statistics to reduce the leakage during error correction, too, as observed on Ref. \cite{BB84Error}, which shows that giving Bob more refined data to be used in the post-processing step can potentially lead to more efficient error-correction and higher key rate.
	
	Here, we propose a simple strategy to alleviate some of the error-correction leakage: Alice and Bob can try to use more than the ZZ matching basis to encode key (e.g. a combination of ZZ, ZX, XZ, XX), since a misalignment angle larger than $\pi/8$ but smaller than $\pi/4$ simply means that the Z basis signal sent are now closer to the X basis to be measured, and the user can swap the data and still generate key. We have performed a simple test simulation in Appendix D, where we observe the effectiveness of such an approach. Nonetheless, as this is not part of the main problem (which focuses on how fine-grained statistics affect the privacy amplification), to simplify the discussion, in the simulations in this paper we will still use a traditional coding strategy of choosing matching bases, i.e. ZZ basis or ZZ and XX bases.
	
	\subsection{Coarse-Graining and Decoy States}
	
	A point worth noting is that, while in principle all observed events can be used as constraints in the SDP problem, there are some situations where some kind of coarse graining is necessary. One such situation occurs if tools like the squashing model \cite{squashing_model} are used. In that case a particular coarse graining is required , depending on the precise measurement set-up.  Another situation is where the analysis of the full fine-grained situation simply takes too much computational resources to calculate the key rate, and coarse graining reduces the problem size.  
	
	For instance, for BB84, a squashing model is often used on Bob's detection system, which corresponds to a coarse graining into $5$ outcomes  that one would expect from a single-photon set-up (vacuum detection, and for qubit detection events). Bob's actual detector patterns, which might also have multi-click events due to incoming multi-photon signals or dark counts, would then need to be coarse-grained correspondingly.. For MDI-QKD, in principle Charlie can use all 16 detector patterns and announce the raw statistics publicly, but to reduce computational complexity, the patterns can be binned into 3 POVMs, corresponding to $\Psi^+$, $\Psi^-, \varnothing$.  Note that such coarse-graining does not affect our ability to characterize the misalignment in the channel (as the cross-basis data between Alice and Bob is still preserved).
	
	When combining coarse-graining and decoy-state analysis we have the freedom to choose between two orders:
	
	(1) apply decoy-state analysis to the raw patterns first (e.g. $4\times16$ times for BB84 and $4\times4\times16$ times for MDI-QKD), and then apply coarse-graining to the estimated single-photon statistics and obtain expectations of the POVMs;
	
	(2) first coarse-grain the raw pattern data obtained from WCP sources to map into the POVM expectations, and then apply decoy states on these smaller number of entries ($4\times5$ entries for BB84 and $4\times4\times3$ entries for MDI-QKD).
	
	The above two orders for performing decoy-state analysis and mapping above are actually theoretically equivalent. The reason is that, in the decoy state analysis, we typically use no  mixing of constraint between different events, so combining variables will not affect the upper and lower bounds.

	\section{Channel Model}

	\begin{figure}[t]
		\includegraphics[scale=0.3]{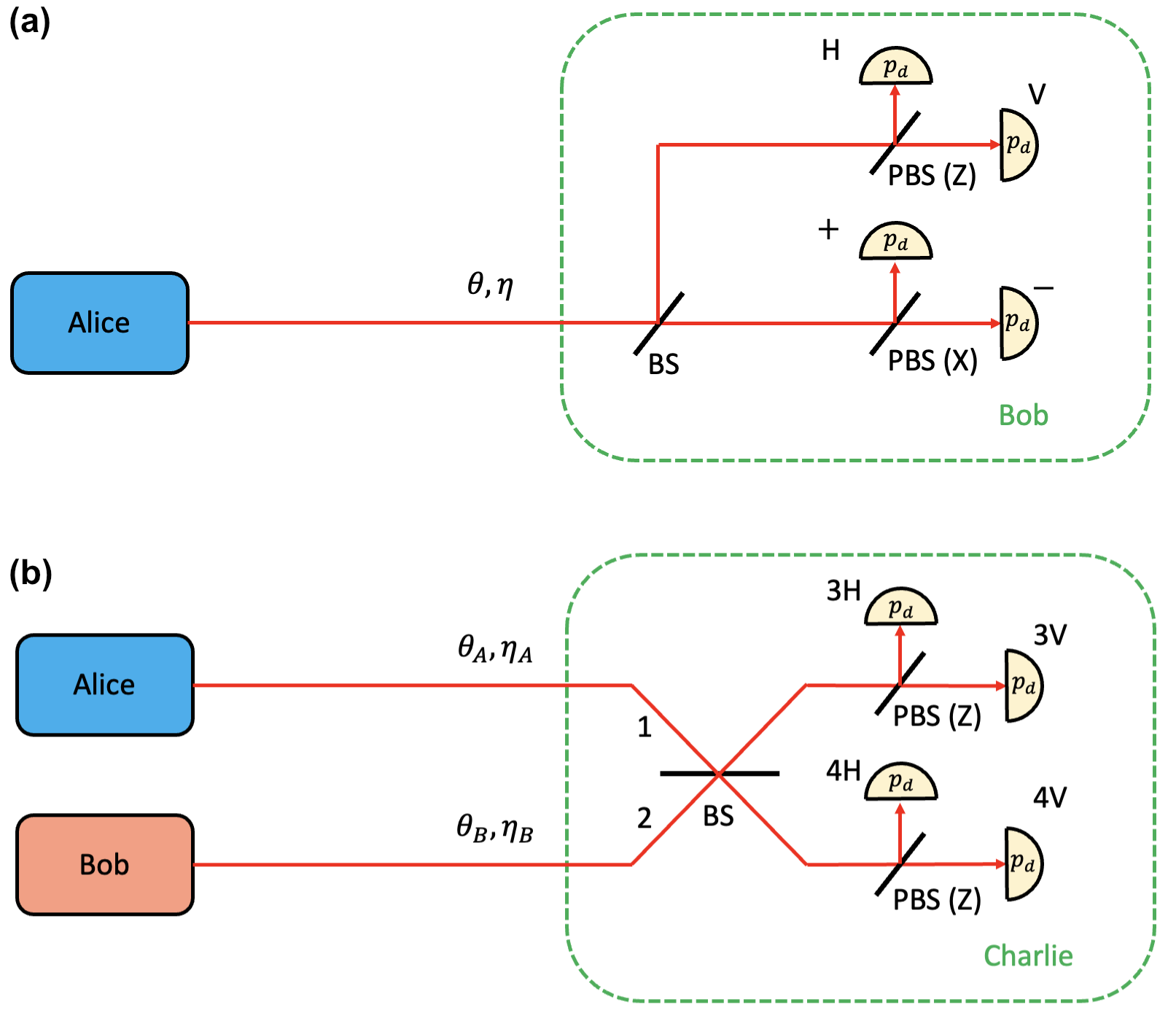}
		\caption{The setup and channel model of BB84 and MDI-QKD. (a) For BB84, Alice's signals go through a rotation $\theta$ along the Z-X plane, and also suffer from channel loss (transmittance $\eta$). Here we show the passive detection scheme for BB84 (while in the active case, the beamsplitter can be replaced by an optical switch). (b) For MDI-QKD, Alice's and Bob's signals each go through a rotation of $\theta_A,\theta_B$ (again, along Z-X plane) with respect to Charlie's measurement basis. Here we assume Charlie always measures in Z basis. The channel model also include Alice's and Bob's losses $\eta_A,\eta_B$. Each detector is assumed to have a dark count rate of $p_d$.}
		\label{fig:QKDSetup}
	\end{figure} 

	\begin{table*}
		
		\begin{tabular}{|c|c|cccc|cccc|}
			\hline
			&&\multicolumn{4}{c|}{Bob's detectors (passive detection)}&\multicolumn{4}{c|}{Bob's detectors (active detection)}\\
			\hline
			&&H&V&+&-&H&V&+&-\\ 
			\hline
			&H & $\sqrt{p_Z}\:cos\:\theta\:\:$&$\sqrt{p_Z}\:sin\:\theta\:\:$&$\sqrt{p_X}\:cos\:\alpha\:\:$&$\sqrt{p_X}\:sin\:\alpha\:\:$
			& $cos\:\theta\:\:$&$sin\:\theta\:\:$&$cos\:\alpha\:\:$&$sin\:\alpha$\\ 
			\multirow{2}{*}{Alice}&V & -$\sqrt{p_Z}\:sin\:\theta\:\:$&$\sqrt{p_Z}\:cos\:\theta\:\:$&$\sqrt{p_X}\:sin\:\alpha\:\:$&$-\sqrt{p_X}\:cos\:\alpha\:\:$
			& $-sin\:\theta\:\:$&$cos\:\theta\:\:$&$sin\:\alpha\:\:$&$-cos\:\alpha$\\ 
			\multirow{2}{*}{sends}&+ & $\sqrt{p_Z}\:sin\:\alpha\:\:$&$\sqrt{p_Z}\:cos\:\alpha\:\:$&$\sqrt{p_X}\:cos\:\theta\:\:$&$-\sqrt{p_X}\:sin\:\theta\:\:$
			& $sin\:\alpha\:\:$&$cos\:\alpha\:\:$&$cos\:\theta\:\:$&$-sin\:\theta$\\ 
			&- & $\sqrt{p_Z}\:cos\:\alpha\:\:$&$-\sqrt{p_Z}\:sin\:\alpha\:\:$&$\sqrt{p_X}\:sin\:\theta\:\:$&$\sqrt{p_X}\:cos\:\theta\:\:$
			& $cos\:\alpha\:\:$&$-sin\:\alpha\:\:$&$sin\:\theta\:\:$&$cos\:\theta$\\ 
			\hline
			
		\end{tabular}
		\caption{Amplitudes arriving at Bob's detectors, for passive and active detection schemes. Here we denote $\alpha = \pi/4 - \theta$. All terms should be additionally multiplied by a factor of $\sqrt{\mu\eta}$, which represents the source and the channel loss. For passive detection scheme, $p_Z,p_X$ are Bob's basis choice probability (implemented in the beamsplitter).}
	\end{table*}
	
	In this section we will describe the channel model we use for simulating the raw statistics that would have been obtained from WCP sources for the BB84 and MDI-QKD protocols (example setups of which are illustrated in fig. \ref{fig:QKDSetup}). These statistics are used in the decoy-state analysis to bound the single photon contribution. Note that all the channel models are only used to simulate the statistics $\gamma_k^{1,U},\gamma_k^{1,L}$, which are independent from the protocol descriptions (i.e. POVMs $\Gamma_k$, and the Kraus operators and key maps). In practice, the raw statistics can also come from an experiment instead.
	
	\subsection{BB84}

	For the channel model, first we need to simulate the statistics coming from the WCP source, before combining these statistics in decoy-state analysis and obtaining the single photon contribution. 
	
	We consider a channel with loss, misalignment, and dark count rate for detectors. Suppose a single photon is sent, the misalignment can be considered as a unitary rotation of angle $\theta$ of the incoming polarization mode. 
	
	In practice, here we are considering an input coherent state of given amplitude, and the channel can be described by the outcome amplitudes arriving at each detector. If we send a signal state with amplitude $\sqrt{\mu}$, and follow the signal through the lossy and misaligned channel and through the linear optics setup at the receiver, the arriving amplitudes (as columns) at the four detectors are listed out in Table III (passive detection).

	Once the amplitude arriving at each detector is known, we can obtain the click probability for a given detector:
	
	\begin{equation}
		\begin{aligned}
			p^{\text{click}}_{j|i} &= 1 - (1-p_d)\times e^{-|\alpha_{j|i}|^2}.\\
		\end{aligned}
	\end{equation}

	\noindent where $\alpha_{j|i}$ is the amplitude for detector $j$ given that Alice prepared state $i$, and $p_d$ is the detector dark count rate. The click probabilities can be further combined into probabilities for each detection pattern from the set of detectors, which can be then mapped via coarse-graining into statistics corresponding to Bob's POVMs. The details of such a mapping can be found in Appendix C.

	\subsection{MDI-QKD}
	
	Now, let us describe the channel model we use for the MDI-QKD protocol with WCP sources. Note that the channel model for MDI-QKD here have been studied in Refs. \cite{MDIPrac, MDIchannel}, and interference of WCP sources have been considered in e.g. \cite{Beamsplitter}.
	
	We consider a channel model as in Fig. \ref{fig:QKDSetup}, where Alice and Bob's signals each have a rotated polarization of $\theta_A,\theta_B$, with respect to Charlie's measurement basis, which is in Z basis only here. The angles $\theta_A,\theta_B$ include the effect of both polarization encoding and polarization misalignment, for instance for input $HV$ the angles would be $(0+\theta_e,\pi/2 + \theta_e)$, where $\theta_e$ is the misalignment angle (we can define misalignment by its induced error rate $e_d=\text{sin}\:\theta_e^2$). The channels each have $\eta_A,\eta_B$ transmittances, and Charlie's detectors have dark count rate $p_d$ for each detector.
	
	As H and V are different modes, we can simply consider the input as having e.g. $\sqrt{\mu_A}\:\text{cos}\:\theta_A$ amplitude in H mode and $\sqrt{\mu_A}\:\text{sin}\:\theta_A$ amplitude in V mode (and similarly for Bob). Before we apply the phase randomization, we can denote the phase difference between the pulses as $\phi$. In this case, the amplitudes at the detectors can be calculated by \cite{Beamsplitter}:
	
	\begin{equation}
	\begin{aligned}
		\alpha_{3H}^\phi = \sqrt{\mu_A \: \eta_A \: \text{cos}\:\theta_A/2} + i\sqrt{\mu_B \: \eta_B \: \text{cos}\:\theta_B/2}  \: e^{i\phi},\\
		\alpha_{4H}^\phi  = i\sqrt{\mu_A \: \eta_A \: \text{cos}\:\theta_A/2} + \sqrt{\mu_B \: \eta_B \: \text{cos}\:\theta_B/2}  \: e^{i\phi},\\
		\alpha_{3V}^\phi  = \sqrt{\mu_A \: \eta_A \: \text{sin}\:\theta_A/2} + i\sqrt{\mu_B \: \eta_B \: \text{sin}\:\theta_B/2}  \: e^{i\phi},\\
		\alpha_{4V}^\phi  = i\sqrt{\mu_A \: \eta_A \: \text{sin}\:\theta_A/2} + \sqrt{\mu_B \: \eta_B \: \text{sin}\:\theta_B/2}  \: e^{i\phi},\\
	\end{aligned}
	\end{equation}

	\noindent where the output ports are denoted $3,4$ and the polarization $H,V$, as shown in Fig. \ref{fig:QKDSetup}. Each detector's clicking probability (including the effect of dark counts) can be written as:
	
	\begin{equation}
		p^{\text{click},\phi}_{k|ij} = 1 - (1-p_d)\times e^{-|\alpha_{k|ij}^\phi|^2}
	\end{equation}
	
	\noindent where $\alpha_{k|ij}^\phi$ is the amplitude at a given detector $k$, for a given set $i,j$ of input from Alice and Bob (representing a combination of $\theta_A,\theta_B$), and conditional to relative phase $\phi$ between incoming signals. Again, it is possible to combine click probabilities for each detector into the probability for each detection pattern $p_{\text{pattern}|ij}^{\phi}$(details are included in Appendix C).
	
	Importantly, for MDI-QKD, (as is observed in Ref. \cite{MDIPrac,MDIchannel,Beamsplitter}), the two incoming coherent states are both phase-randomized, which means the relative phase $\phi$ between two incoming signals must be integrated between $0$ to $2\pi$:
	
	\begin{equation}
		p_{\text{pattern}|ij} = {1/{2\pi}} \: \int_{0}^{2\pi}  \: p_{\text{pattern}|ij}^\phi  \: d\phi.
	\end{equation}

	Afterwards, the detection pattern data can be coarse-grained into statistics that correspond to the POVMs (in this case, Charlie's three classical announcements corresponding to two Bell states and failure event).

	Up to here we have obtained the sets of raw statistics, which can be used in the decoy-state analysis. Again, as discussed in Sec. II. E, the order of coarse-graining and decoy-state statistics is arbitrary. The details of the coarse-graining and some example data can be found in Appendix C.

	\section{Simulated Key Rate Results}
	
	\begin{figure}[h]
		\includegraphics[scale=0.14]{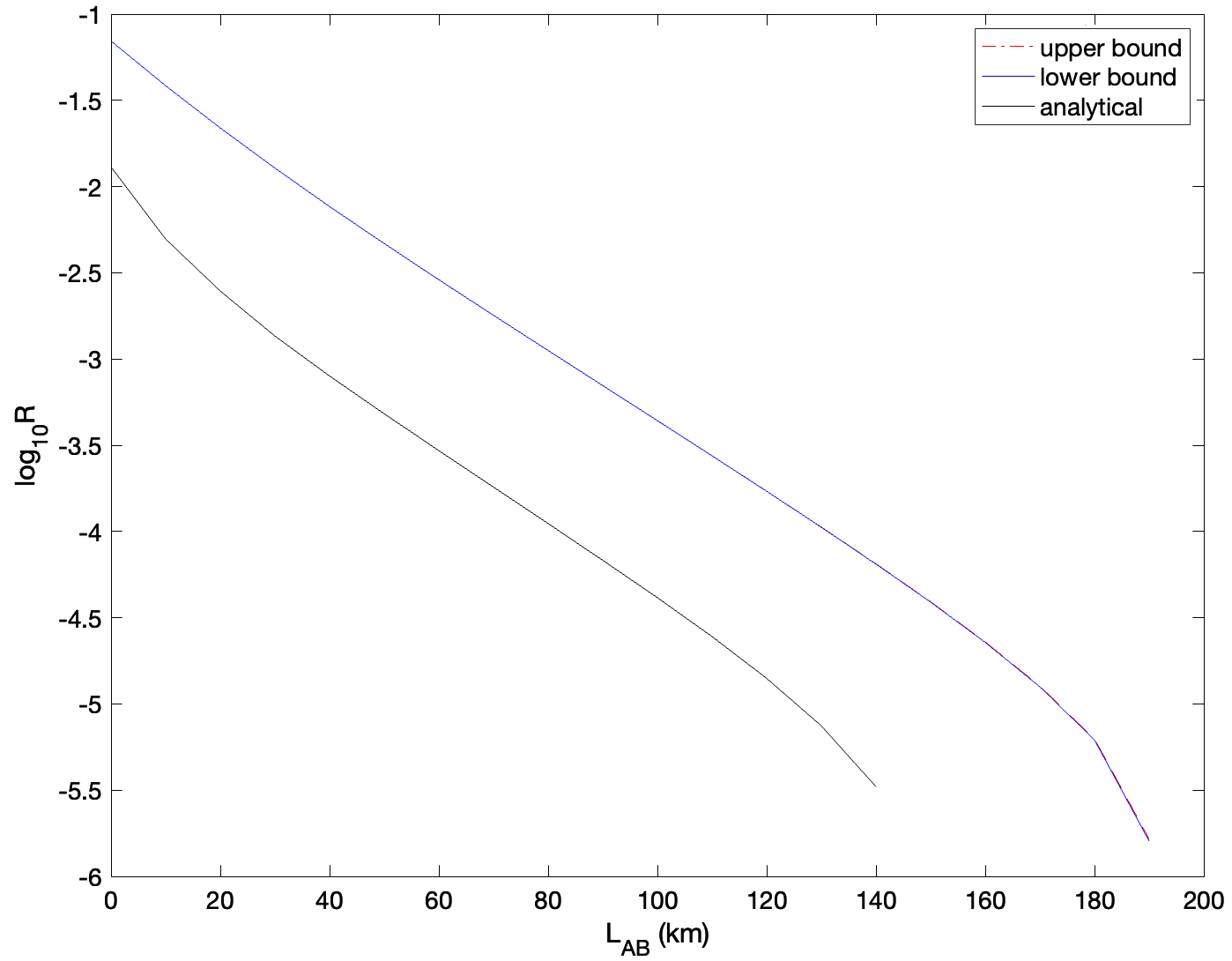}
		\caption{Simulation results for decoy-state BB84 key rate versus distance, generated for our numerical approach with fine-grained statistics versus prior art analytical approach in Ref. \cite{decoy_practical}. The plots are generated at a considerable level of misalignment between Alice and Bob of $\theta=0.3$ (corresponding to $e_d=8.7\%$), and the strong decoy intensity $\mu$ is optimized (ranging from approximately 0.2-0.4), and $\nu,\omega$ are fixed as shown in Table IV. The new approach is shown to have consistently higher key rate than the analytical approach. Also, the upper/lower bounds are shown to be very tight (mostly overlapping in the plot).}
		\label{fig:rate_BB84}
	\end{figure} 

	\begin{figure}[h]
		\includegraphics[scale=0.14]{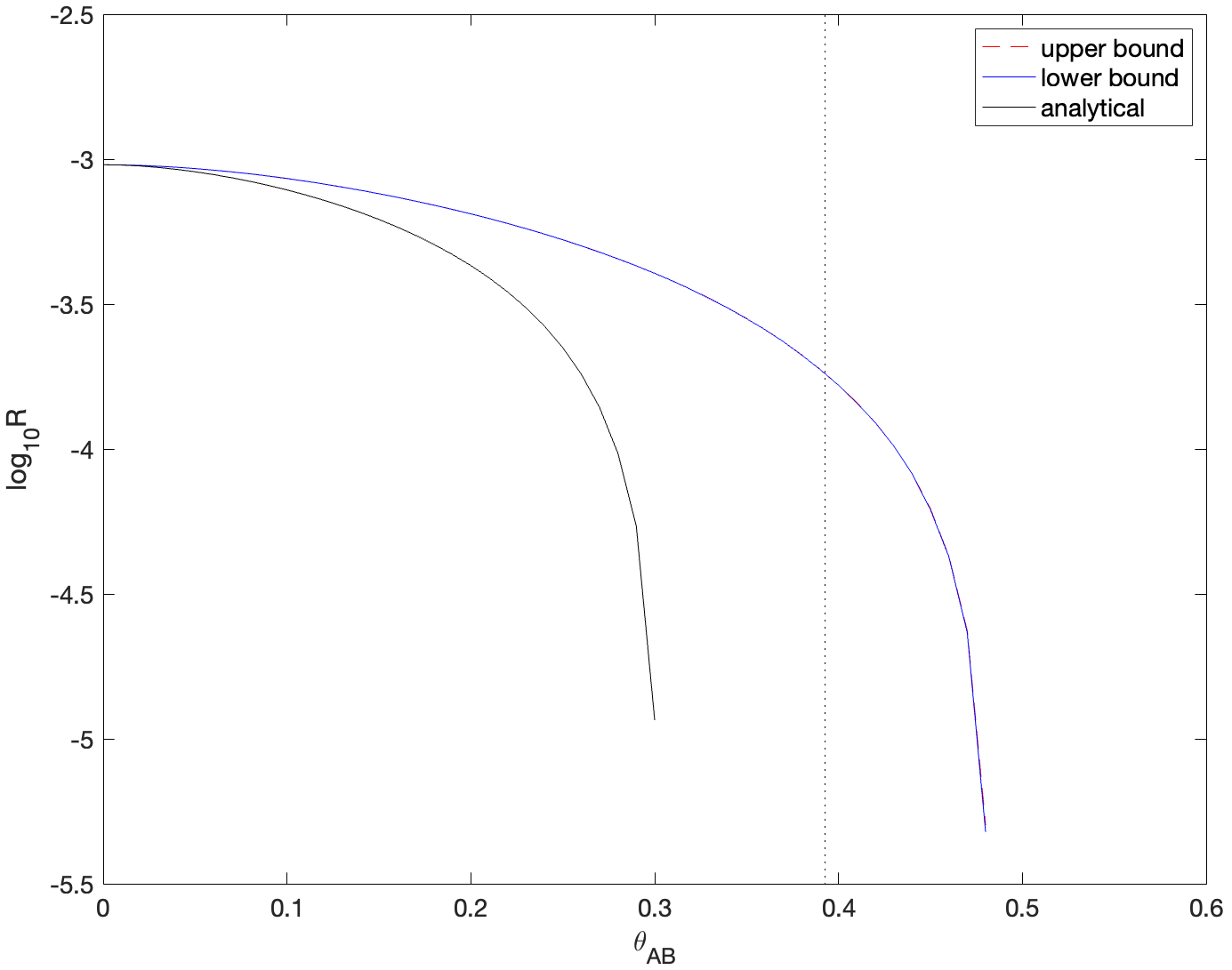}
		\caption{Simulation results for decoy-state BB84 for key rate versus various misalignment angles at $L_{AB}=100km$, $p_d=10^{-6}$, generated for our numerical approach with fine-grained statistics versus analytical approach \cite{decoy_practical}. As can be seen, the maximum tolerable angle has greatly increased, e.g. for $R=10^{-5}$ the tolerable misalignment (note that $e_d=sin^2\:\theta$) increases from approximately 8.7\% to almost 21\%. Note that for the numerical approach with fine-grained data, key rate will still decrease with misalignment, but it is only due to error correction leakage and no longer from Eve's correlation with the quantum signals. Note that, when the misalignment angle is larger than $\pi/8$ (marked as dotted vertical line) and if Alice and Bob are able to know this, it is possible for them to simply use a different basis combination for key generation, since the arriving signals originally sent in Z basis are now closer to the X basis. This would affect both fine-grained and coarse-grained analytical approaches. More details on better error-correction strategies can be found in Appendix C.5.}
		\label{fig:angle_BB84}
	\end{figure}

	\begin{figure}[h]
		\includegraphics[scale=0.14]{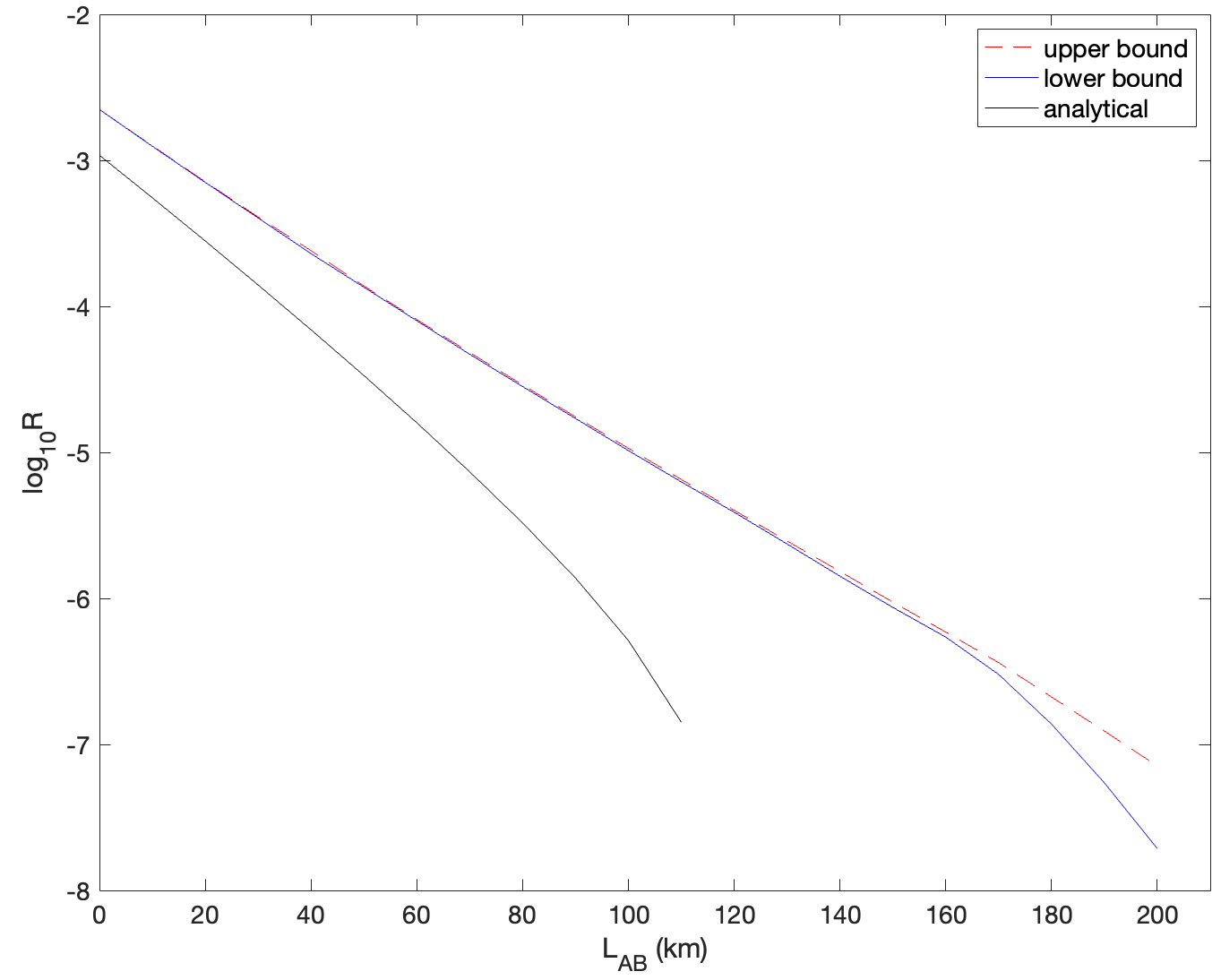}
		\caption{Simulation results for decoy-state MDI-QKD key rate versus total distance, generated for our numerical approach with fine-grained statistics versus analytical approach in Ref. \cite{MDI}. Alice's and Bob's misalignment is set to a considerable level of $\theta_A=0.15,\theta_B=-0.15$ (which is equivalent to about 8.7\% misalignment between Alice and Bob). For MDI-QKD, solving for the key rate is significantly more computationally intensive, so here we have used fixed $\mu_A=\mu_B=0.25$ instead of performing an optimization. We can see that we consistently have higher key rate across all distances. Note that at long distances, due to the small values of observed statistics, numerical noises start to show for the SDP solver, which makes the lower bound less tight.}
		\label{fig:rate_MDI}
	\end{figure} 

	\begin{figure}[h]
		\includegraphics[scale=0.14]{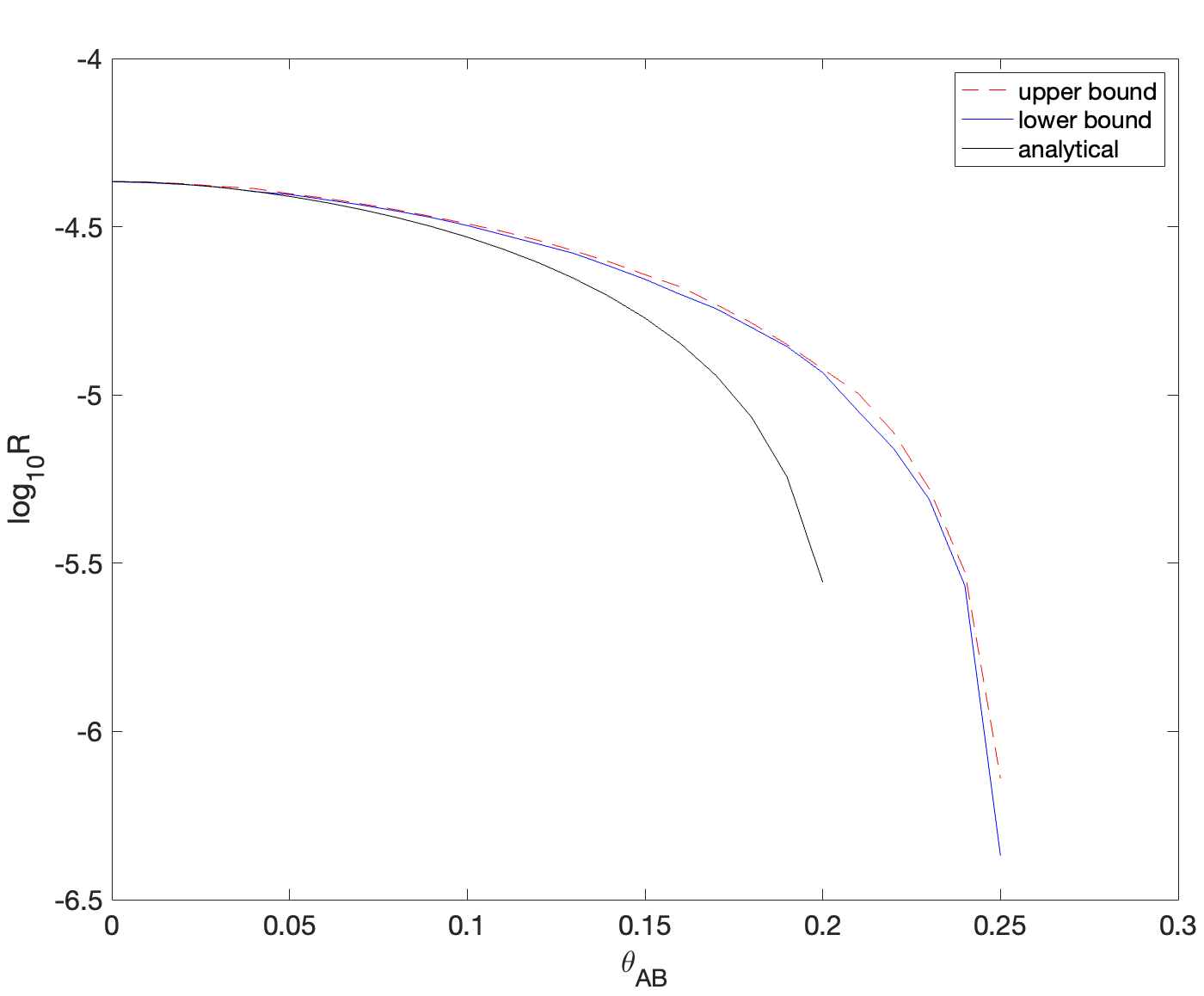}
		\caption{Simulation results for decoy-state MDI-QKD key rate versus various misalignment angles at $L_{AB}=100km$, $p_d=10^{-6}$, generated for our numerical approach with fine-grained statistics versus analytical approach \cite{MDI}. Here we are scanning $\theta_A$ while keeping $\theta_B=0$. As can be seen, the maximum tolerable angle has considerably increased, e.g. for $R=10^{-5.5}$ the tolerable misalignment increases from approximately 4\% to 5.6\%. Note that for the numerical approach, key rate will still decrease with misalignment, but it is only due to error correction leakage and no longer from privacy amplification.}
		\label{fig:angle_MDI}
	\end{figure}

	\begin{table}
		
		\begin{tabular}{ccccc}
			\hline\hline
			$p_d$  & $\alpha$ & $f$ & $\nu$ & $\omega$\\
			\hline
			$1\times 10^{-6}$ & 0.2dB/km & 1 & 0.02 & 0.001 \\
			\hline\hline
			
		\end{tabular}
	\caption{Simulation parameters including dark count rate, fibre loss per km, and error-correction efficiency. For simplicity, detector efficiency is included into the total loss $\eta$, and error-correction efficiency is assumed to be perfect (equals 1). Here we have used a set of fixed weak+vacuum decoy intensities $\nu,\omega$, along with a strong decoy intensity $\mu$ (which can be either fixed or optimized).}
	\end{table}

	In this section we have performed simulations of the key rate versus distance and key rate versus misalignment angle, for BB84 and MDI-QKD respectively, using the parameters from Table IV.
	
	In Fig. \ref{fig:rate_BB84}, we plot the key rate of the numerical approach with fine-grained statistics and compare it with using coarse-graining (which uses analytical key rate formula), in the presence of strong misalignment $e_d=8.7\%$. As we can see, using the full-statistics demonstrates a considerable advantage over using traditional analytical approaches for decoy-state BB84, both providing higher key rate and tolerating higher levels of misalignment angle. We also plot the key rate versus misalignment angle in Fig. \ref{fig:angle_BB84}, and we can see that at e.g. $L_{AB}=100km$, decoy-state BB84 can now tolerate up to $0.48$ rad of misalignment angle (equivalent to $e_d=21\%$). In comparison, using coarse-graining and analytical approach, one can only tolerate around only $0.3$ rad of misalignment angle, which is only 60\% that of the case with fine-graining.
	
	We also plot the key rate for MDI-QKD against both distance and misalignment angle for the numerical approach with fine-grained statistics, and compare it to the analytical approach with coarse-graining. From Figs. \ref{fig:rate_MDI}, \ref{fig:angle_MDI}, we can see similar behavior as the BB84.Again we can see that here the fine-graining provides higher resilience against misalignment, allowing larger tolerable misalignment angles and enabling longer maximum communication distance when misalignment is large.
	
	\section{Conclusion and Discussions}
	
	In this work we have utilized a well-established numerical framework and applied it to decoy-state BB84 and MDI-QKD, which extends the practical scenarios the framework can be applied to. We show that decoy-state analysis can be incorporated into the framework as a preprocessing \textit{wrapper} to generate pseudo-statistics used for key rate calculation, which in principle allows any protocol that can be described in the framework to be uplifted to a decoy-state based protocol. Additionally, we show that using full fine-grained statistics including cross-basis events, we can gain higher performance and acquire reference-frame-independence in the privacy amplification and gain higher key rate. Importantly, this allows us to directly apply the new analysis to existing systems and even existing data, and readily gain a better key rate, which can simplify experimental design as it reduces the need for manual alignment, and can even potentially extend the maximum distance the QKD protocol can achieve.
	
	Note that for this approach, there are two important limitations so far: (1) Similar to RFI-QKD, here we have to assume that the misalignment angle is only slowly drifting and not quickly changing with time (such that a constant rotation angle can be assumed); (2) the analysis in this work works only for the asymptotic (infinite data) regime, and combing decoy states with the finite-size analysis  \cite{numerical_finite} in the numerical framework will be the subject of future works. 

	\section{Acknowledgments}
	
	We would like to thank in particular Jie Lin for constructive suggestions and contribution to the code. We would also like to thanks Shlok A Nahar, Twesh Upadhyaya, Eli Bourassa, and Hoi-Kwong Lo for helpful discussions. We acknowledge the support from Huawei, NSERC Collaborative Research and Development (CRD) Program and Discovery Grants Program, and Compute Canada. Institute for Quantum Computing (IQC) is supported by Innovation, Science and Economic Development (ISED) Canada.

\appendix

\section{Protocol Descriptions}

\subsection{BB84}

The BB84 description is the same as in the model of Ref. \cite{numerical2}. Note that to model a threshold detector, we should consider a qubit squashing model, and assign five POVM elements (four BB84 states and a failure case representing more than one clicks or no click). In the description we should specify the POVM operators, Kraus operators, and key maps).

In the protocol, Alice encodes in a local four-dimensional qudit entangled with the flying qubit sent to Bob:
\begin{equation}
	\ket{\psi}_{AA'}=(\ket{0}\ket{H}+\ket{1}\ket{V}+\ket{2}\ket{+}+\ket{3}\ket{-})/2
\end{equation}

Bob measures with a qubit squashing model \cite{squashing_model}, which maps all input into a three-dimensional POVM (single-photon and vacuum subspace)

\begin{equation}
	\begin{aligned}
	P_1^B &= p_Z \: \begin{pmatrix}
		0 & 0 & 0 \\
		0 & 1 & 0 \\
		0 & 0 & 0 \\
	\end{pmatrix},\\
	P_2^B &= p_Z \: \begin{pmatrix}
		0 & 0 & 0 \\
		0 & 0 & 0 \\
		0 & 0 & 1 \\
	\end{pmatrix},\\
	P_3^B &= {1\over 2} \: p_X \: \begin{pmatrix}
		0 & 0 & 0 \\
		0 & 1 & 1 \\
		0 & 1 & 1 \\
	\end{pmatrix},\\
	P_4^B &= {1\over 2} \: p_X \: \begin{pmatrix}
		0 & 0 & 0 \\
		0 & 1 & -1 \\
		0 & -1 & 1 \\
	\end{pmatrix},\\
	P_5^B &= \begin{pmatrix}
		1 & 0 & 0 \\
		0 & 0 & 0 \\
		0 & 0 & 0 \\
	\end{pmatrix}.\\
	\end{aligned}
\end{equation}

The overall POVM acting on the density matrix between Alice and Bob should be the kronecker product of above five operators and Alice's four local qubit bases. 

There are also some additional deterministic constraints in the form of:

\begin{equation}
\ket{i} \bra{j}_A \otimes \mathbb{I}_{dim_B},
\end{equation} 

\noindent where $i,j$ are each combination of states for system A. The constraints trace over Alice's local qubit and characterize our knowledge of the source, i.e. we know exactly that Alice sent $\ket{\psi}_{AA'}$.

Alice and Bob perform sifting to keep the cases where they used the same basis. Consider four systems: Alice's local register storing the basis, Alice's qudit, Bob's qubit, and the register storing the announcement. The Kraus operators are:

\begin{equation}
	\begin{aligned}
		K_Z&=\left[\begin{pmatrix}1 \\ 0\end{pmatrix} \: \otimes \: \begin{pmatrix}1&&&\\&0&&\\&&0&\\&&&0\end{pmatrix} + \begin{pmatrix}0 \\ 1\end{pmatrix} \: \otimes \: \begin{pmatrix}0&&&\\&1&&\\&&0&\\&&&0\end{pmatrix}\right]  \\
		& \: \otimes \: \sqrt{p_Z} \: \begin{pmatrix}1&&\\&1&\\&&1\end{pmatrix}  \: \otimes \:  \begin{pmatrix}1 \\ 0\end{pmatrix}\\
		K_X&=\left[\begin{pmatrix}1 \\ 0\end{pmatrix} \: \otimes \: \begin{pmatrix}0&&&\\&0&&\\&&1&\\&&&0\end{pmatrix} + \begin{pmatrix}0 \\ 1\end{pmatrix} \: \otimes \: \begin{pmatrix}0&&&\\&0&&\\&&0&\\&&&1\end{pmatrix}\right] \\
		&  \: \otimes \: \sqrt{p_X} \: \begin{pmatrix}1&&\\&1&\\&&1\\\end{pmatrix}  \: \otimes \:  \begin{pmatrix}0 \\ 1\end{pmatrix}\\
	\end{aligned}
\end{equation}

\noindent while the key maps are:

\begin{equation}
	\begin{aligned}
		Z_1=\begin{pmatrix}1 & 0 \\ 0 & 0\end{pmatrix} \:  \otimes \:  \mathbb{I}_{\text{dim}_A \times \text{dim}_B \times 2}\\
		Z_2=\begin{pmatrix}0 & 0 \\ 0 & 1\end{pmatrix}  \: \otimes \:  \mathbb{I}_{\text{dim}_A \times \text{dim}_B \times 2}
	\end{aligned}
\end{equation}

\noindent where the dimensions are $\text{dim}_A=4,\text{dim}_B=3$ here, and $ \mathbb{I}_x$ is an identity matrix of $x\times x$.

\subsection{MDI-QKD Description}

The MDI-QKD description is the same as in the model of Ref. \cite{numerical2}. In this setup, Alice (Bob) keeps a local qubit system A (B), and send an entangled system a (b). The two system a,b are received by Charles who performs a bell state measurement (BSM) and publicly announces the results.

In MDI-QKD, Charles can only distinguish two out of the four Bell states $\ket{\Phi^\pm}=(\ket{HV}\pm\ket{VH})/\sqrt{2}$. More than two clicks, wrong click patterns, and no clicks are all considered inconclusive. There are therefore three POVM operators for Charles:

\begin{equation}
	\begin{aligned}
		P^C_1&=\ket{\Phi^+}_{ab}\bra{\Phi^+}_{ab}, \\
		 P^C_2&=\ket{\Phi^-}_{ab}\bra{\Phi^-}_{ab},\\
		  P^C_3&=1-P^C_1-P^C_2.\\
	\end{aligned}
\end{equation}

Here we can construct the experimental observables (which are the Kronecker's product between Alice's and Bob's local measurements and Charles' BSM):

\begin{equation}
	\Gamma_{ijk} = P^A_i  \: \otimes \:  P^B_j  \: \otimes \:  P^C_k.
\end{equation} 

Again, we also include the deterministic constraints (partial traces over Alice's and Bob's systems only)

\begin{equation}
	\ket{i} \bra{j}_A \otimes \ket{k} \bra{l}_B \otimes \mathbb{I}_{dim_C}
\end{equation} 

\noindent to characterize our knowledge of the source. If we perform decoy-state and limit ourselves to only the single-photon subspace, the expectation values for these deterministic constraints would correspond to that of a single-photon source.

Next, the Kraus operators are:

\begin{equation}
	\begin{aligned}
		K_Z&=\left[\begin{pmatrix}1 \\ 0\end{pmatrix} \: \otimes \: \begin{pmatrix}1&&&\\&0&&\\&&0&\\&&&0\end{pmatrix} + \begin{pmatrix}0 \\ 1\end{pmatrix} \: \otimes \: \begin{pmatrix}0&&&\\&1&&\\&&0&\\&&&0\end{pmatrix}\right]  \\
		&  \: \otimes \:  \begin{pmatrix}1&&& \\ &1&& \\ &&0& \\ &&&0 \end{pmatrix} \:  \otimes \:  \begin{pmatrix}1 &  &  \\  & 1 &  \\  & & 0\end{pmatrix} \: \otimes \:  \begin{pmatrix}1 \\ 0\end{pmatrix},\\
		K_X&=\left[\begin{pmatrix}1 \\ 0\end{pmatrix} \: \otimes \: \begin{pmatrix}0&&&\\&0&&\\&&1&\\&&&0\end{pmatrix} + \begin{pmatrix}0 \\ 1\end{pmatrix} \: \otimes \: \begin{pmatrix}0&&&\\&0&&\\&&0&\\&&&1\end{pmatrix}\right] \\
		&  \: \otimes \:  \begin{pmatrix}0&&& \\ &0&& \\ &&1& \\ &&&1 \end{pmatrix}  \: \otimes \:  \begin{pmatrix}1 &  &  \\  & 1 &  \\  & & 0\end{pmatrix} \: \otimes \:  \begin{pmatrix}0 \\ 1\end{pmatrix},\\
	\end{aligned}
\end{equation}

\noindent while the key maps are:

\begin{equation}
	\begin{aligned}
		Z_1=\begin{pmatrix}1 & 0 \\ 0 & 0\end{pmatrix}  \: \otimes \:  \mathbb{I}_{\text{dim}_A \times \text{dim}_B \times \text{dim}_C \times 2},\\
		Z_2=\begin{pmatrix}0 & 0 \\ 0 & 1\end{pmatrix}  \: \otimes \:  \mathbb{I}_{\text{dim}_A \times \text{dim}_B \times \text{dim}_C \times 2},
	\end{aligned}
\end{equation}

\noindent here $\text{dim}_A=4,\text{dim}_B=4,\text{dim}_C=3$, where the first two are dimensions of Alice's and Bob's local qubits, and $\text{dim}_C$ is the number of Charles' POVM outcomes.

\section{Details on Implementing Decoy-State Analysis with Linear Programming}

This section is a brief recapitulation of the techniques for performing decoy-state analysis, e.g. in Refs. \cite{decoy_practical, MDIPrac}, or the review paper Ref. \cite{decoyreview}. As mentioned in the main text, the problem of solving for upper/lower bounds on single photon contribution can be formulated as a \textit{linear programming} problem. Denote the observable of interest as $\gamma$, and the single photon statistics as $\gamma_1$ while overall observed statistics from the WCP source is $\gamma_\mu$. Then we can write:

\begin{equation}
	\begin{aligned}
		\gamma_{\mu_i} &= \sum p_{\mu_i}(n)  \: \gamma_n.
	\end{aligned}
\end{equation}

\noindent $p_{\mu_i}(n)$ is the Poissonian distribution:

\begin{equation}
	p_{\mu_i}(n) = {{\mu_i}^n \over n!}  \: e^{-{\mu_i}} .
\end{equation}

The above set variables $\gamma_n$ are bounded by a set of linear equations (the number of equations corresponds to the number of decoy intensities $\{\mu_i\}$). The set of equations constitutes a linear programming problem, and any of the observables such as $\gamma_1$ can be upper-bounded and lower-bounded to obtain $\gamma_1^L,\gamma_1^U$. Note that the key observation here is that an eavesdropper cannot tell apart the pulses coming from decoy states, hence $\gamma_n$ is the same variable in all equations, independent of the intensity $\mu_i$.

Decoy-state analysis for MDI-QKD is similar, as described in \cite{MDI}. The difference is that there are two sources, hence two Poissonian distributions:

\begin{equation}
	\begin{aligned}
		p_{\mu_{Ai}}(n_A) = {{\mu_{Ai}}^{n_A} \over {n_A}!} \: e^{-{\mu_{Ai}}}, \\
		p_{\mu_{Bj}}(n_B) = {{\mu_{Bj}}^{n_B} \over {n_B}!} \: e^{-{\mu_{Bj}}}, 
	\end{aligned}
\end{equation}

\noindent the linear constraints then become:

\begin{equation}
	\begin{aligned}
		\gamma_{\mu_{Ai},\mu_{Bj}} &= \sum_{n_A} \: \sum_{n_B} \:  p_{\mu_{Ai}}(n_A)  \: p_{\mu_{Bj}}(n_B)  \: \gamma_{n_A,n_B} \\
	\end{aligned}
\end{equation}

\noindent where Alice and Bob respectively choose their decoy intensities $\mu_{Ai},\mu_{Bj}$ from the available sets. For instance, if Alice and Bob each chooses from three intensities, here we will have 9 equations from the decoy state data, which can be used in the linear program. The terms of interest is the single-photon pair contribution $Y_{1,1}$.

When there are infinite levels of decoys (i.e. infinite number of equations), we can perfectly bound any variable. In practice, however, usually three decoys (say $\mu,\nu,\omega$) for Alice (BB84) or for each of Alice and Bob (MDI-QKD) is sufficient to tightly bound the single photon contribution $Y_1$ or $Y_{1,1}$. Also, as we cannot model infinite number of variables in the actual computation, a cutoff (say $S=10$) needs to be implemented, and all higher-photon-number terms are upper bounded by 1 and lower-bounded by 0. For instance, the linear program constraints for BB84 can be rewritten as:

\begin{equation}
	\begin{aligned}
		\gamma_{\mu_i} &\leq \sum_{n\leq S}  \: p_{\mu}(n)  \: \gamma_n  + (1-\sum_{n\leq S}  \: p_{\mu}(n)),\\
		\gamma_{\mu_i} &\geq \sum_{n\leq S}  \: p_{\mu}(n)  \: \gamma_n.\\
	\end{aligned}
\end{equation}

\noindent (which is Eq. 10 in the main text). 

Similar applies to MDI-QKD, where photon numbers for Alice and Bob satisfy some cutoff region. For instance if we allow them to each independently satisfy the cutoff condition, the constraints will be:

\begin{equation}
	\begin{aligned}
		\gamma_{\mu_i} &\leq \sum_{n_A\leq S}\sum_{n_B\leq S}  \:p_{\mu_{Ai}}(n_A)  \: p_{\mu_{Bj}}(n_B)  \: \gamma_{n_A,n_B}  \\
		&+ (1-\sum_{n_A\leq S}\sum_{n_B\leq S} \: p_{\mu_{Ai}}(n_A)  \: p_{\mu_{Bj}}(n_B)),\\
		\gamma_{\mu_i} &\geq \sum_{n\leq S}  \: p_{\mu_{Ai}}(n_A)  \: p_{\mu_{Bj}}(n_B)  \: \gamma_{n_A,n_B}.\\
	\end{aligned}
\end{equation}

One formulated, the problem can be passed to an external linear program solver (such as Gurobi) to solve efficiently. Alternative to the linear program, it is also possible to derive an analytical bound for the single photon statistics, such as in Refs. \cite{MDIPrac,MDI7int}.

\section{Processing of Detection Data for BB84 and MDI-QKD}

In this section we describe how to simulate the raw detection data and how it can be coarse-grained into the POVM expectations used for the key rate calculation.

\subsection{Simulating Detection Data for BB84}

In the main text Section III, we derived the amplitudes of coherent light arriving at each detector. For each detector, the click probability is:

\begin{equation}
	\begin{aligned}
		p^{\text{click}}_{j|i} &= 1 - (1-p_d)\times e^{-|\alpha_{j|i}|^2}, \\
	\end{aligned}
\end{equation}

There are a total of four detectors, but with possibility of multi-counts, leading to a total of 16 possible detection patterns. The probabilities for each detection pattern $b_1b_2b_3b_4$ ($b_j=0$ if no click, and $b_j=1$ if a click) can be generated by

\begin{equation}
	\begin{aligned}
		p_{b_1b_2b_3b_4|i} &= \Pi_{j=1,2,3,4}  \: \{\overline{b_j} + p^{\text{click}}_{j|i}  \: (-1)^{\overline{b_j}}\}
	\end{aligned}
\end{equation}

\noindent where $\overline{b_j}$ is the bit flip of $b_j$ (such that the term in curly bracket is $p_{\text{click}}$ for $b_j=1$, and $1-p_{\text{click}}$ for $b_j=0$). As an example, for the pattern $1100$ (H and V detectors double-clicked and the two other detectors didn't click), the pattern probability $p_{\text{1100}|i} =p^\text{click}_{1|i} \: p^\text{click}_{2|i}  \: (1-p^\text{click}_{3|i} ) \: (1-p^\text{click}_{4|i})$.\\

Note that in the above model we have assumed a passive detection setup (where basis choice is performed by a beamsplitter), which is also what we will use in the simulations in this paper. For reference, for active detection, the amplitudes arriving at the detectors for each given signal state can also be found in Table III, where there is no beam-splitter that splits the intensity. The basis choice is actively performed, and the expectation value is the sum of the expectations from the two choices.

\begin{equation}
	\begin{aligned}
		p^{\text{click}}_{\text{basis},j|i} &= b_{\text{basis},j} \: [1 - (1-p_d)\times e^{-|\alpha_{j|i}|^2}], \\
		p_{b_1b_2b_3b_4|i}^Z &= \Pi_{j=1,2,3,4}  \: \{\overline{b_j} + p^{\text{click}}_{Z,j|i} (-1)^{\overline{b_j}}\}, \\
		p_{b_1b_2b_3b_4|i}^X &= \Pi_{j=1,2,3,4}  \: \{\overline{b_j} + p^{\text{click}}_{X,j|i} (-1)^{\overline{b_j}}\}, \\
		p_{b_1b_2b_3b_4|i} &= p_Z  \: p_{b_1b_2b_3b_4|i}^Z  +p_X  \: p_{b_1b_2b_3b_4|i}^X,
	\end{aligned}
\end{equation}

\noindent where again $\overline{b_j}$ is the bit flip of $b_j$, and the logical bit for basis choice $b_{Z,j}$ is 1 for $j=1,2$ (and 0 otherwise), and $b_{X,j}$ is 1 for $j=3,4$ (and 0 otherwise), which represents the activation/deactivation of detectors depending on the basis. The terms inside curly brackets behave the same way as in the passive detection scenario, taking value of $p^{\text{click}}_{\text{basis},j|i}$ if the bit in the detection pattern $b_j=1$, and $(1-p^{\text{click}}_{\text{basis},j|i})$ otherwise. The only difference is that an additional logical bit $b_{\text{basis},j}$ turns off all X basis detectors (hence suppressing all patterns where X basis detectors click) when measuring in Z basis, and vice versa, such that there are no cross-clicks on two bases simultaneously. The final click probability is the weighted average between X and Z basis statistics.

After we calculate all $4\times 16$ entries of $p_{b_1b_2b_3b_4|i}$, we have all the raw detection statistics for the given decoy intensity setting, which we can denote as $P_{raw,\mu}$. The full simulation/experiment involves collecting all sets of decoys. For instance, if Alice uses three decoy intensities $\{\mu,\nu,\omega\}$, the statistics would be $P_{raw,\mu},P_{raw,\nu},P_{raw,\omega}$. 

The above statistics can be used in the decoy-state analysis to obtain single-photon contributions. As mentioned in Sec. II. E, we can perform the coarse-graining and decoy-state analysis in arbitrary order. The details of the coarse-graining process and example data can be found in the next subsection.

\subsection{Coarse-Graining Map for BB84}

\begin{figure}[h]
	\includegraphics[scale=0.16]{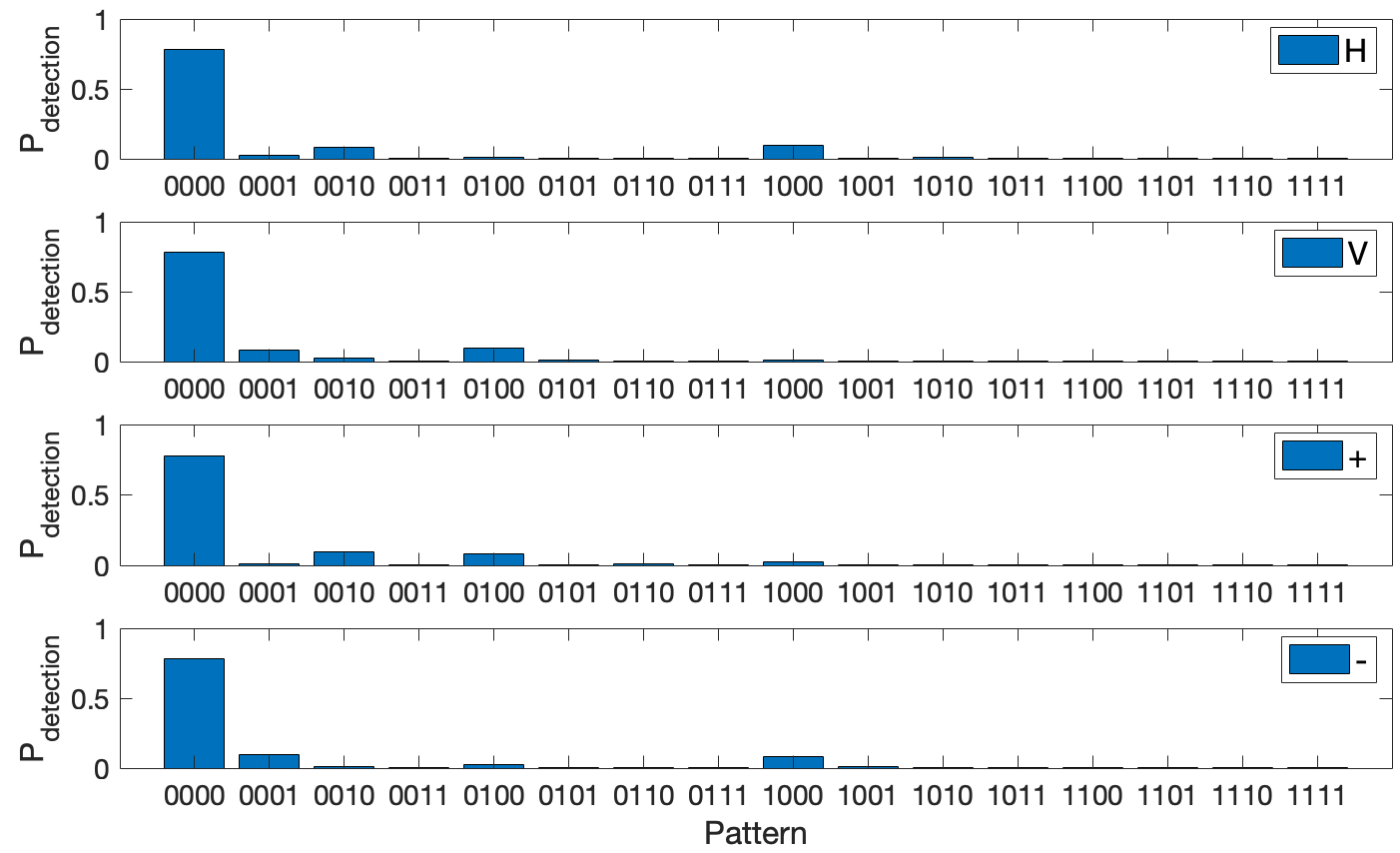}
	\caption{Example BB84 raw detection patterns, generated at $\eta=1$  and  $\theta=0.3, p_d=10^{-6}$, for the strong decoy states $\mu=0.25$. The mapping $M$ is then applied to the raw patterns to obtain a 5-POVM qubit squashing model. Note that after the mapping, the obtained $4\times 5$ matrix is the statistics for a given decoy state (e.g. $\mu$), for which further decoy state analysis is needed to obtain the actual single photon contribution. Here we only show one set of data for $\mu$, but e.g. two more sets from $\nu,\omega$ are also needed for decoy-state analysis.}
	\label{fig:pattern_BB84}
\end{figure} 

\begin{figure}[h]
	\includegraphics[scale=0.14]{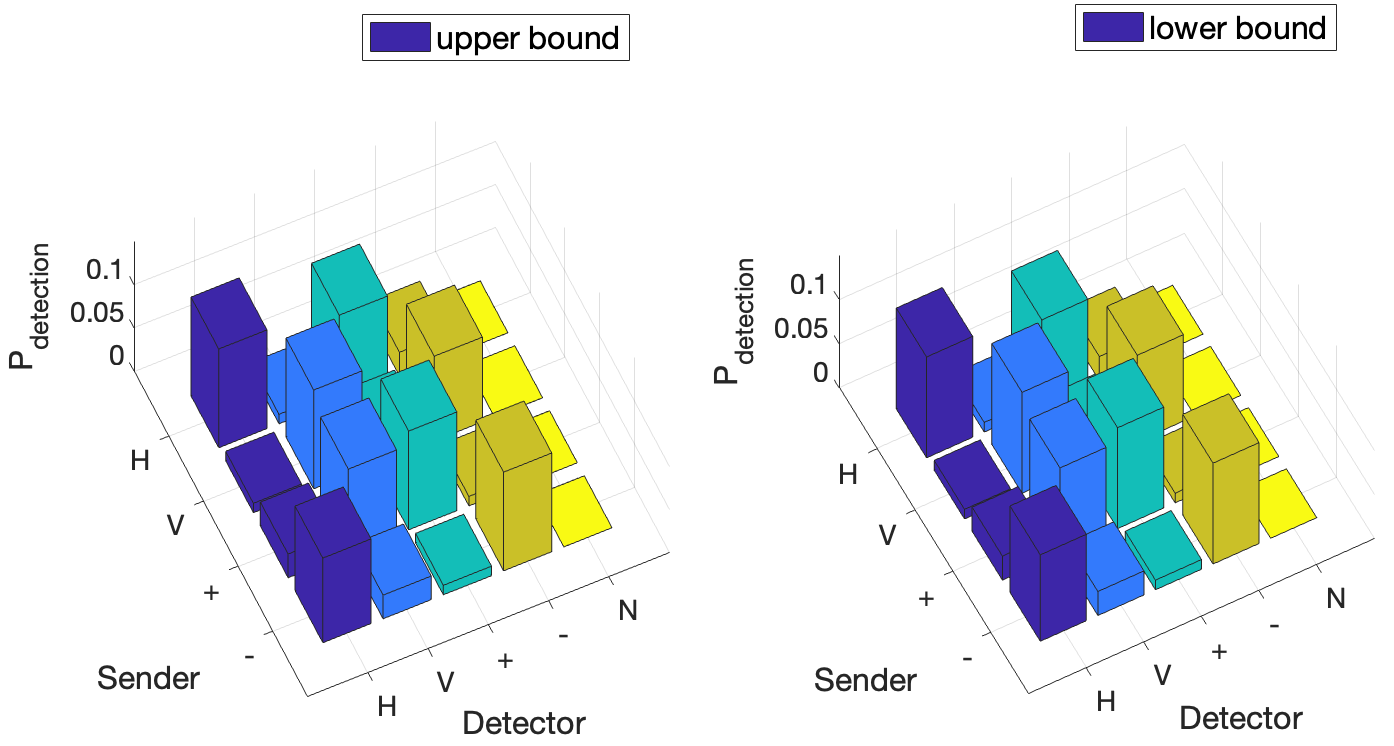}
	\caption{Example BB84 upper and lower bounds for single photon statistics obtained from decoy-state analysis and after the mapping, generated at $\eta=1$  and  $\theta=0.3, p_d=10^{-6}$. Note that the decoy-state analysis can be done before or after the mapping (i.e. on each term of the $4\times 16$ raw data, or on each term of the $4\times 5$ after mapping). Here for BB84 we choose to do decoy states on the $4\times 5$ decoy statistics after the mapping is applied to the raw statistics above in e.g. Fig. \ref{fig:pattern_BB84}. After the decoy state analysis, the upper and lower bounds for the single-photon statistics, including the cross-basis data, can be then directly used by the numerical framework, which feeds these statistics into the SDP solver as constraints to solve for the worst density matrix $\rho$ that lower bounds the key rate.}
	\label{fig:pattern3D_BB84}
\end{figure} 

	For BB84, Alice sends for signal states, and Bob detects with four detectors. The raw data containing click/no-click events would constitute 16 different patterns for the four detectors. A total of $4\times16$ entries for each intensity setting chosen can be recorded from the simulation/experiment. This matrix can be mapped back to a 5-entry qubit-based statistics corresponding to the POVMs, by applying a mapping matrix $M$. The mapping rules are: double clicks in the same basis is assigned randomly to a bit, while clicks simultaneously in different bases are discarded for passive detection (for active detection there are no such clicks, since only detectors in one basis are activated at a time).
	
	\begin{equation}
		\begin{aligned}
			M_{H} &= [0,0,0,0,0,0,0,0,1,0,0,0,0.5,0,0,0] \\
			M_{V} &= [0,0,0,0,1,0,0,0,0,0,0,0,0.5,0,0,0] \\
			M_{+} &= [0,0,1,0.5,0,0,0,0,0,0,0,0,0,0,0,0] \\
			M_{-} &= [0,1,0,0.5,0,0,0,0,0,0,0,0,0,0,0,0] \\
			M_{\varnothing} &= [1,0,0,0,0,1,1,1,0,1,1,1,0,1,1,1] \\
			M&=[M_H^T,M_V^T,M_+^T,M_-^T,M_\varnothing^T]\\
		\end{aligned}
	\end{equation}
	
	\noindent where the elements are binary-encoded (i.e. $j=8b_1 + 4b_2 + 2b_3 + b_4$). The single-photon statistics can be obtained by simply applying the mapping $M$ on the raw statistics $p^{\text{full},\mu}$:
	
	\begin{equation}
		P_{\mu}= P_{\text{raw},\mu} \times M.
	\end{equation}
	
	We can apply the mapping to the raw detection data for each intensity setting, to obtain a $4\times5$ matrix for each intensity setting, for instance three sets of $P_{\mu},P_{\nu},P_{\omega}$, if three decoy settings $\{\mu,\nu,\omega\}$ are used. 
	
	Then, for each entry $\gamma_{k}$ in the matrix ($k$ corresponds to an entry in the matrix where a given signal is sent by Alice and a given event is detected by Bob, for instance $HH$), we have three entries $\gamma_{\mu,k},\gamma_{\nu,k},\gamma_{\omega,k}$ from the matrices $P_{\mu},P_{\nu},P_{\omega}$, with which we can perform decoy-state analysis, and obtain the upper and lower bounds for the single photon component $\gamma_{1,k}^L,\gamma_{1,k}^U$ for the event $k$. Performing decoy-state analysis 20 times for each event, we can obtain two matrices $P_1^L,P_1^U$, which are the full single-photon statistics (corresponding to Table I) that can be fed into the SDP solver.
	
	Example illustrations for the raw detection pattern statistics and the single-photon statistics obtained from decoy states can be seen in Figs. \ref{fig:pattern_BB84}, \ref{fig:pattern3D_BB84}.
	
	\subsection{Simulation of Detection Data for MDI-QKD}
	
	From the main text Sec. III we have already obtained the amplitudes arriving at each detector, which are functions of the phase difference $\phi$ between incoming signals. Our goal is to convert them to a detection pattern probability (which is still a function of $\phi$), and lastly integrate it over $\phi \in [0,2\pi)$ to obtain the statistics for two independently phase-randomized sources.
	
	Again, given the amplitude and the phase difference $\phi$, the click probability for a given detector is 
	
	\begin{equation}
		p^{\text{click},\phi}_{k|ij} = 1 - (1-p_d)\times e^{-|\alpha_{k|ij}^\phi|^2}
	\end{equation}
	
	The probability of observing a given detector pattern $b_1b_2b_3b_4$ (0 or 1 representing no click versus click), just like for the four detectors for BB84 in the previous subsections, is:
	
	\begin{equation}
		\begin{aligned}
			p_{b_1b_2b_3b_4|ij}^\phi&= \prod_{k=1,2,3,4} \: \{\overline{b_k} + p^{\text{click}}_{k|ij}  \: (-1)^{\overline{b_k}}\} \\
		\end{aligned}
	\end{equation}
	
	\noindent where again we denote $\overline{b_k}$ as the bit flip of $b_k$ (whose effect is to calculate $p_{\text{click}}$ for $b_k=1$, and $1-p_{\text{click}}$ for $b_k=0$). 
	
	Note that the detector pattern probability above $p_{b_1b_2b_3b_4|ij}^\phi$ is for a given phase difference between incoming pulses, $\phi$. To take into account the phase randomization and calculate the average statistics, $p_{b_1b_2b_3b_4|ij}^\phi$ needs to be integrated over all $\phi$
	
	\begin{equation}
		p_{b_1b_2b_3b_4|ij} = {1/{2\pi}} \: \int_{0}^{2\pi}  \: p_{b_1b_2b_3b_4|ij}^\phi  \: d\phi.
	\end{equation}
	
	\noindent At this point we have a set of $p_{b_1b_2b_3b_4|ij}$ for each input $i,j$ and each pair of given intensities $\mu_A,\mu_B$, which is a matrix of $4\times4\times16$ patterns in total. We can denote such a matrix as $P_{raw,\mu_{A}\mu_{B}}$. For instance, if Alice and Bob each use a set of three decoy intensities $\{\mu,\nu,\omega\}$, there would be a total of 9 sets of statistics for each combination of $\mu_{Ai},\mu_{Bi}$:
	
	\begin{equation}
		\begin{aligned}
			&P_{\text{raw},\mu\mu}, \: P_{\text{raw},\mu\nu}, \: P_{\text{raw},\mu\omega},\\
			&P_{\text{raw},\nu\mu}, \: P_{\text{raw},\nu\nu}, \: P_{\text{raw},\nu\omega},\\
			&P_{\text{raw},\omega\mu}, \: P_{\text{raw},\omega\nu}, \: P_{\text{raw},\omega\omega}.
		\end{aligned}
	\end{equation}

	With these statistics, we can apply decoy-state and optional coarse-graining to obtain the bounds on single-photon statistics, which can be further used to bound the key rate. We will discuss the process in more detail in the next subsection.

	\subsection{Coarse-Graining Map for MDI-QKD}
	
	For MDI-QKD, Alice and Bob each sends four signal states, and Charlie uses four detectors. The raw data containing click/no-click events contain $4\times4\times16$ entries for each intensity setting. The mapping rules are simple: only the four patterns corresponding to Bell states $\Psi^+,\Psi^-$ will be kept, and all other patterns are discarded:
	
	\begin{figure}[h]
		\includegraphics[scale=0.16]{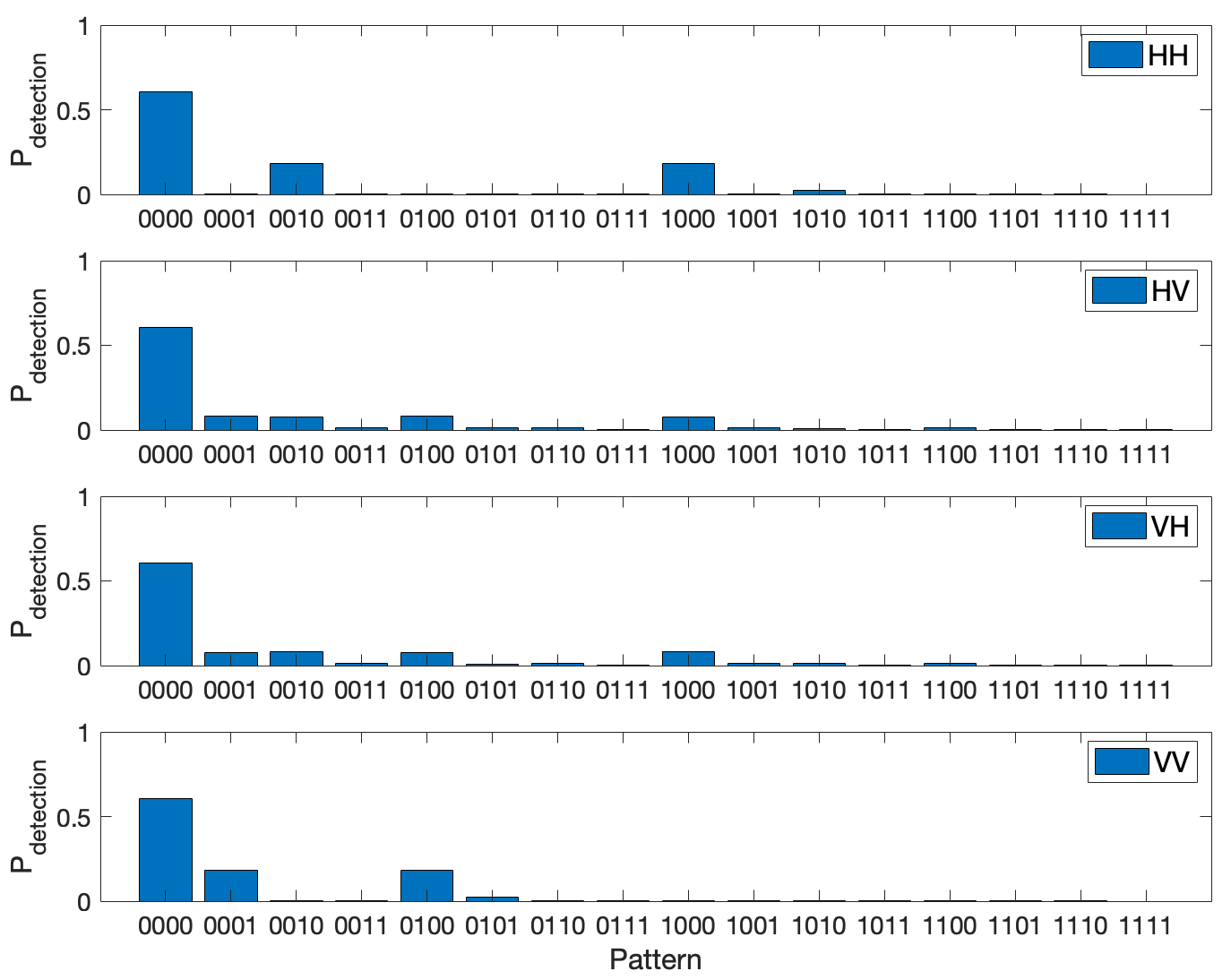}
		\caption{Example MDI-QKD raw detection patterns (due to limited space, here only the Z basis signals are shown, while X basis and cross-basis clicks are not shown), generated at $\eta=1$  and  $\theta_A=0.2,\theta_B=0$, for the strong decoy states $\mu_A=\mu_B=0.25$. For each pattern and for each Alice and Bob's sent state (e.g. HH state, pattern 0001) a decoy state analysis is required to estimate the upper and lower bounds from single photon contributions. After obtaining the pattern probabilities for single photons using decoy state analysis, the mappings are then applied to the patterns to obtain Bell state event statistics.}
		\label{fig:pattern}
	\end{figure} 
	
	\begin{figure}[h]
		\includegraphics[scale=0.15]{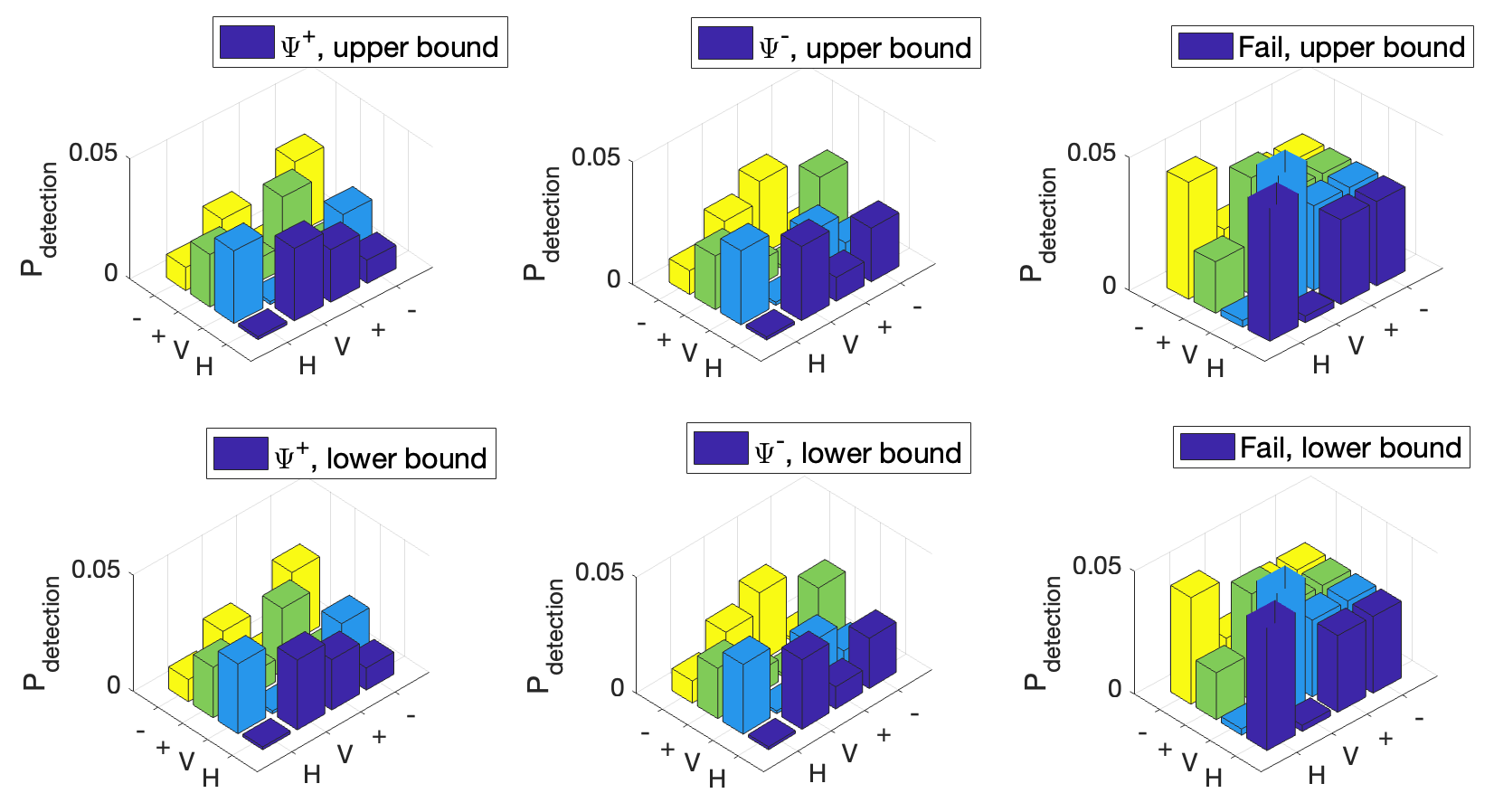}
		\caption{Example MDI-QKD upper and lower bounds for single photon statistics obtained from decoy-state analysis and after the mapping, generated at $\eta=1$ and  $\theta_A=0.2,\theta_B=0$. The X and Y axes are Alice and Bob's sent signals. Note that the cross-basis data are not necessarily symmetric here if $\theta_A \neq \theta_B$. Again, after decoy state analysis, the single-photon statistics including the cross-basis information can be fed into the SDP solver as constraints to solve for the worst density matrix $\rho$ that lower bounds the key rate.}
		\label{fig:SPpattern}
	\end{figure} 

\begin{equation}
	\begin{aligned}
		M_{\Psi^-} &= [0,0,0,0,0,0,1,0,0,1,0,0,0,0,0,0] \\
		M_{\Psi^+} &= [0,0,0,1,0,0,0,0,0,0,0,0,1,0,0,0] \\
		M_{\varnothing} &= 1_{1\times 16} - M_{\Psi^+} -M_{\Psi^-} \\
		M&=[M_{\Psi^-}^T,M_{\Psi^+}^T,M_{\varnothing}^T]\\
	\end{aligned}
\end{equation}

\noindent where $M_{\Psi^-}$ represents the patterns 1001 and 0110, and $M_{\Psi^+} $ represents the patterns 1100 and 0011, while all other patterns are binned into failure events, represented by $M_{\varnothing}$. After defining the maps, the patterns can be acquired by applying the maps (used as constraints for the actual solver).

\begin{equation}
	P_{\mu_A\mu_B} =P_{raw,\mu_A\mu_B} \times M.
\end{equation}

If Alice and Bob each uses three intensity settings, there will be a set of 9 $P_{\mu_A\mu_B}$. They can perform decoy-state analysis on each entry of $P_{\mu_A\mu_B}$ to obtain the bounds on the single photon statistics $P_1$. Example sets of such statistics before and after the mapping can be seen in Figs. \ref{fig:pattern}, \ref{fig:SPpattern}. The single photon statistics $P_1$ (or more precisely, its upper and lower bounds) can be directly fed into the numerical solver as the constraints $\{\gamma_k\}$ to lower bound the secure key rate. \\

Alternatively, one can choose to perform decoy-state first on $P_{raw,\mu_A\mu_B}$ to obtain $P_{raw,1}$, where the mapping process becomes: 
\begin{equation}
	P_{1} =P_{raw,1} \times M.
\end{equation}

\subsection{Using Cross-Basis Data in Key Generation}

\begin{figure}[h]
\includegraphics[scale=0.14]{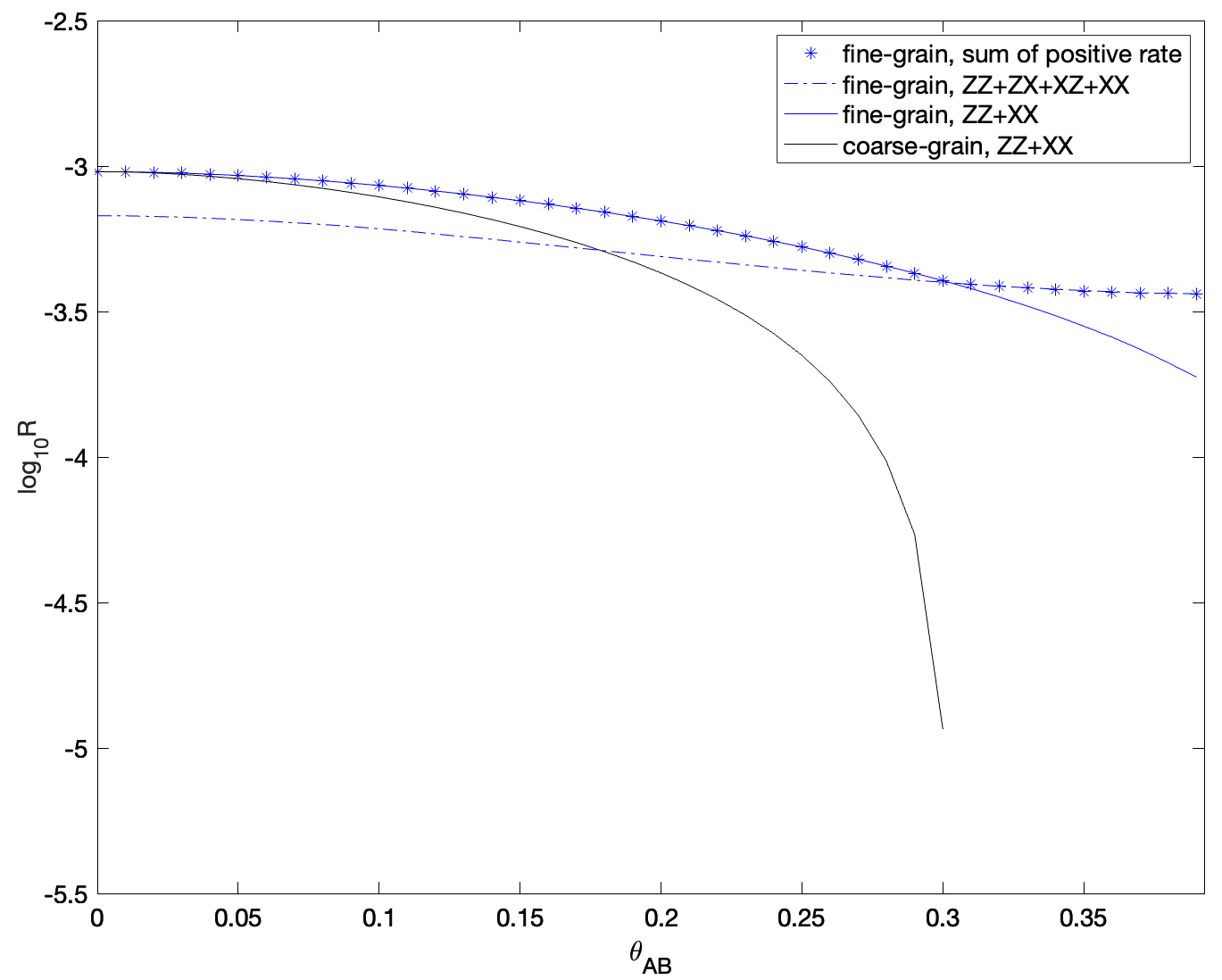}
\caption{Comparison of using all basis combinations simultaneously for encoding (dot-dash blue line) versus using ZZ and XX (solid blue line), for decoy-state BB84 key rate plotted against misalignment angle between Alice and Bob. The coarse-grain key rate (solid black line) is also included, generated analytically. We also include the case where key rates for ZZ, ZX, XZ, XX bases are independently generated and only combinations with positive key rates are taken (shown in blue stars), which as shown here happen to be the joint upper bound for key rates using ZZ+XX or key rate using all basis combinations simultaneously. The parameters used for simulation are identical to Fig. \ref{fig:angle_BB84}, i.e. from Table IV, and at 100km distance between Alice and Bob. The plot is only generated up to $\theta=\pi/8$, because beyond that point, the key rate is symmetrical for strategies using all bases, while strategies using only matching bases ZZ+XX can be simply switched to ZX+XZ if the users find out that $\theta>\pi/8$.}
\label{fig:basis_compare}
\end{figure} 

In this subsection we propose a better choice of bases to generate key in the presence of a large misalignment angle.

Here we look again at the key rate formula:

\begin{equation}
	R = \text{min}_{\rho \in S}  \: f(\rho) - p_{\text{pass}} \times \text{leak}^{\text{EC}}_{\text{obs}}.
\end{equation}

\noindent As we have described in the main text, the fine-grained statistics is able to provide more information to effectively bound $\rho$, and cancel out the effect of misalignment in the privacy amplification term $\text{min}_{\rho \in S} f(\rho)$ of the key rate. However, the leakage $\text{leak}^{\text{EC}}_{\text{obs}}$ is a classical observed value that cannot be reduced by such methods.

Now, let us first consider a worst-case scenario for BB84, where the misalignment angle is $\pi/4$: the signals sent in Z basis by Alice but rotated by the channel would be equivalent to ones sent in X basis to begin with. If Bob measures in Z basis, he would get completely random results and cannot generate key. However, if Bob instead measures in X basis, he would get perfect statistics, if no other sources of error are included. This means that, if the misalignment angle is between 0 and $\pi/4$ degrees, then at least two sets - or even all four sets - among the normally sifted bases (sending-receiving in Z-Z and X-X) and the cross bases (sending-receiving in Z-X and X-Z) would be able to have a low QBER and generate key. The worse case scenario here becomes $\pi/8$ instead, at which point both sets of bases would have 15\% QBER (still enough to generate key).

We can then rigorously describe it in our framework. The Kraus operators become:

\begin{equation}
	\begin{aligned}
		K_{ZZ}&=\left[\begin{pmatrix}1 \\ 0\end{pmatrix} \: \otimes \: \begin{pmatrix}1&&&\\&0&&\\&&0&\\&&&0\end{pmatrix} + \begin{pmatrix}0 \\ 1\end{pmatrix} \: \otimes \: \begin{pmatrix}0&&&\\&1&&\\&&0&\\&&&0\end{pmatrix}\right]  \\
		& \: \otimes \: \sqrt{p_Z} \: \begin{pmatrix}1&&\\&1&\\&&1\end{pmatrix}  \: \otimes \:  \begin{pmatrix}1 \\ 0 \\0 \\0\end{pmatrix},\\
		K_{ZX}&=\left[\begin{pmatrix}1 \\ 0\end{pmatrix} \: \otimes \: \begin{pmatrix}1&&&\\&0&&\\&&0&\\&&&0\end{pmatrix} + \begin{pmatrix}0 \\ 1\end{pmatrix} \: \otimes \: \begin{pmatrix}0&&&\\&1&&\\&&0&\\&&&0\end{pmatrix}\right]  \\
		& \: \otimes \: \sqrt{p_X} \: \begin{pmatrix}1&&\\&1&\\&&1\end{pmatrix}  \: \otimes \:  \begin{pmatrix}0 \\ 1 \\ 0 \\ 0\end{pmatrix},\\
		K_{XZ}&=\left[\begin{pmatrix}1 \\ 0\end{pmatrix} \: \otimes \: \begin{pmatrix}0&&&\\&0&&\\&&1&\\&&&0\end{pmatrix} + \begin{pmatrix}0 \\ 1\end{pmatrix} \: \otimes \: \begin{pmatrix}0&&&\\&0&&\\&&0&\\&&&1\end{pmatrix}\right]  \\
		& \: \otimes \: \sqrt{p_Z} \: \begin{pmatrix}1&&\\&1&\\&&1\end{pmatrix}  \: \otimes \:  \begin{pmatrix}0 \\ 0 \\ 1 \\ 0\end{pmatrix},\\
		K_{XX}&=\left[\begin{pmatrix}1 \\ 0\end{pmatrix} \: \otimes \: \begin{pmatrix}0&&&\\&0&&\\&&1&\\&&&0\end{pmatrix} + \begin{pmatrix}0 \\ 1\end{pmatrix} \: \otimes \: \begin{pmatrix}0&&&\\&0&&\\&&0&\\&&&1\end{pmatrix}\right] \\
		&  \: \otimes \: \sqrt{p_X} \: \begin{pmatrix}1&&\\&1&\\&&1\end{pmatrix}  \: \otimes \:  \begin{pmatrix}0 \\ 0 \\ 0 \\ 1\end{pmatrix},\\
	\end{aligned}
\end{equation}

\noindent where the systems are respectively Alice's register, Alice's local qubit, Bob's POVMs corresponding to a particular basis, and Alice and Bob's announcement registers. The key maps are:

\begin{equation}
	\begin{aligned}
		Z_1=\begin{pmatrix}1 & 0 \\ 0 & 0\end{pmatrix}  \: \otimes \:  \mathbb{I}_{\text{dim}_A \times \text{dim}_B \times 4},\\
		Z_2=\begin{pmatrix}0 & 0 \\ 0 & 1\end{pmatrix}  \: \otimes \:  \mathbb{I}_{\text{dim}_A \times \text{dim}_B \times 4}.
	\end{aligned}
\end{equation}

\noindent The leakage term can be written as:

\begin{equation}
	\begin{aligned}
	\text{leak}^{\text{EC}}_{\text{obs}} &= Q_{ZZ} \: h_2(E_{ZZ}) +  Q_{XZ} \: h_2(E_{XZ}) \\
	&+  Q_{ZX} \: h_2(E_{ZX}) +  Q_{XX} \: h_2(E_{XX})\\
	\end{aligned}
\end{equation}

\noindent where $h_2(x) = - x  \: \text{log}_2 \: (x) - (1-x) \:  \text{log}_2 \:  (1-x)$.\\

Recall that $f(\rho)$ is calculated by:

\begin{equation}
	\begin{aligned}
		f(\rho) &= D(\mathcal{G}(\rho) || \mathcal{Z(\mathcal{G}(\rho))} ), \\
		\mathcal{G}(\rho) &= \sum_i  \: K_i  \: \rho  \: K_i^\dagger, \\
		\mathcal{Z(\mathcal{G}(\rho))} &= \sum_j  \: Z_j \:  \mathcal{G}(\rho) \:  Z_j .\\
	\end{aligned}
\end{equation}

As there are two copies of identical Kraus operators (except the public announcement register), we can get a higher  $f(\rho)$ for the same $\rho$. In the ideal case (e.g. no misalignment and loss), $f(\rho)$ would return twice the value. Of course, this comes at the cost of higher leakage term.

We perform a simple simulation for decoy-state BB84 in Fig. \ref{fig:basis_compare} as a demonstration. As can be seen, using all four combinations of basis simultaneously can generate higher key in some situations with large misalignment angles, while at low misalignment angles it might be better to only use ZZ and XX, because signals with ZX and XZ bases combinations would have large QBER when misalignment angle is small, resulting in large penalties from error-correction leakage.

Moreover, an alternative strategy, which is potentially better, is to simply perform privacy amplification and error-correction independently on each of the portions of signals with ZZ, ZX, XZ, XX basis combinations, and dynamically use only the combinations with positive key rates.

\begin{equation}
	\begin{aligned}
 		R =\: &\text{max}(0,R_{ZZ}) + \text{max}(0,R_{ZX})\\
 		+\:&\text{max}(0,R_{XZ})+\text{max}(0,R_{XX})
	\end{aligned}
\end{equation}

\noindent where $R_{b_i b_j}$ is defined as the key rate obtained by performing privacy amplification and error-correction only on the signals where Alice sends in basis $b_i$ and Bob measures in basis $b_j$. The Kraus operator would be $K_{b_i b_j}$ only, and the leakage would be $Q_{b_ib_j}h_2(E_{b_ib_j})$ only.

Note that while we demonstration this for BB84, in principle MDI-QKD protocols can utilize such a setting to reduce error-correction leakage when misalignment angle is large.

Potentially, there could also be better strategies than the ones we propose here. For instance, users may even directly calculate the entropy based on the $4\times4$ matrix utilizing the statistics of all input combinations of Alice and Bob (instead of choosing a specific pair of bases). Such an approach will be the subject of future studies.

\newpage

\end{document}